# Alcohol additives to enhance ammonia-methane combustion efficiency and reduce emissions: A reactive force field analysis

Amirali Shateri, Zhiyin Yang, Jianfei Xie*, Nasser Sherkat
School of Engineering, University of Derby, DE22 3AW, UK

**Abstract**
Exploring the impact of alcohol additives on combustion and pyrolysis of ammonia/methane is of great importance in the pursuit of sustainable energy technologies. This work employs Reactive Force Field (ReaxFF) molecular dynamics (MD) simulations to investigate the underlying mechanism of how ethanol and methanol additives affect reaction pathways, NOx emissions and bond energy characteristics in ammonia-methane pyrolysis and combustion processes. It shows that adding alcohols altered NOx formation pathways, reducing the diversity of NOx and shifting the equilibrium toward simpler NOx such as NO and NO2. At 2,000 K, alcohol blends, particularly methanol, demonstrated a notable reduction in NO2 formation. At 3,000 K, both ethanol and methanol suppressed NO production, but the influence of methanol was stronger. Nitric acid production, HNO3, was present at lower temperatures but became negligible at higher temperatures because of the thermal breakdown of the higher-order NOx. These trends confirm that alcohol additives realize a probable role in moderating NOx emissions and stabilizing reaction pathways. The pyrolysis in modified reaction pathways, which facilitated the decomposition of ammonia and methane in these blends, affected the formation of intermediate species, leading to the reduction in peak emissions. In addition, methanol and ethanol showed significant impacts on hydrogen bond energies of the mixture, especially important building block radicals encouraging higher complexity pathways. By leveraging a computationally robust and scalable methodology, this study not only advances a fundamental understanding of alcohol-enhanced ammonia/methane combustion but also informs strategies to optimize these mixtures for practical use in modern propulsion systems.

## 1. Introduction

The global drive for clean energy has increased interest in alternative fuels, and ammonia ($NH_3$) seems like a natural candidate for internal combustion, aviation fuel and gas turbines. Ammonia, historically used as an agricultural fertilizer, has garnered industrial interest as a low-carbon energy carrier due to its high energy density. The focus on ammonia as a fuel stems from the urgent need to decarbonize energy generation and consumption as a strategy to mitigate climate change. But ammonia's low burning speed and instability when burnt by itself required finding additives to improve its combustion [1-2]. Langella [3] identified these challenges and highlighted the potential of combustion enhancers, such as alcohols, to mitigate these problems. Recent advancements in zero-emission aviation fuel technology have identified ammonia-methane blends as a promising candidate, with research focusing on their molecular interactions, combustion dynamics and emission behavior. For instance, Xu et al. [4] demonstrated that adding ammonia to hydrocarbon fuel systems significantly affects soot formation and the growth of polycyclic aromatic hydrocarbons (PAHs). More precisely, ammonia was seen to retard the growth of large carbon-containing species in $C_2H_4/O_2$ systems through new paths involving HCN addition. Verified through the assistance of quantum chemical calculations [5], such an inhibiting effect on PAHs growth reflects that ammonia, besides being a carbon-free alternative fuel, would also have effects on preventing certain pollutant formations during combustion. Investigations into high-temperature and high-pressure conditions of ammonia/methane ($NH_3/CH_4$) combustion have revealed detailed reaction mechanisms and pathways [6]. These studies have shown that such conditions as high temperatures and pressures significantly affect ammonia consumption and the formation of nitrogen oxides (NOx); pressure, in particular, complicates reaction pathways owing to increased molecular and atomic collisions. More emphasis shall be put in the identification and characterization of reaction intermediates at high pressures associated with new elementary reactions. This chemistry underscores the complex nature of ammonia-methane combustion while highlighting opportunities for further contributions to its understanding and optimization for practical applications, including net-zero aviation [7]. Notably, NOx emissions were found to be significantly lower for rich fuel-air mixtures, primarily due to NHi (i stands for the number of hydron atoms) reaction pathways facilitating NO consumption in the presence of excess NH3. Huang et al. [8] evaluated whether increasing the ammonia content of methanol-ammonia fuel mixtures changed the engine performance and emissions. Their simulation results found that increasing the ratio of ammonia to substitution caused the average effective pressure, brake power and brake torque to increase. Intriguingly, fuel economy was also a major gain at high speeds, with lower brake specific fuel consumption and greater brake thermal efficiency.

*Corresponding author: j.xie@derby.ac.uk (J.X.)

Moreover, recent technological progress on alternative fuels has focused on zero emission options, and ammonia-methane combustion is one such candidate [9-12]. However, due to the inherent problems with blending ammonia-methane, researchers have proposed novel approaches to modify its burning properties. These include the use of electric fields [13-15], ultrasound waves [16], and alcohol encapsulants such as methanol and ethanol [17,18]. Additives of alcohol in the burning process have proven promising in controlling exhaust gases. Recent research suggests that these additives reduce potentially harmful pollutants typical of fossil fuel combustion−hydrocarbons (HC), carbon monoxide (CO), and nitrogen oxides (NOx). Besides, blending alcohol with conventional fuels may lead to an improvement in brake thermal efficiency. This is because of the higher oxygen content and octane ratings in alcohol fuels, which result in better combustion processes. This aligns with emerging sustainable fuel strategies that emphasize oxygenated components, such as volatile fatty acid–derived paraffins, for cleaner and more efficient combustion, especially in aviation fuels [19]. Zhang et al. [20] investigated the effects of methanol and ethanol on the pyrolysis of n-decane using ReaxFF simulations supported by the experimental verification. Their results indicated that both alcohols significantly influenced the pyrolysis process, altering reaction pathways and product distributions. Independently, Yu et al. [21] investigated the effect of methanol on kerogen pyrolysis using ReaxFF molecular dynamics. Their work demonstrated that adding methanol changed the thermal decomposition profile of kerogen, which in turn affected the yield of the major products produced during pyrolysis. Together, these studies have demonstrated the exceptional potential of alcohol as additives in reshaping hydrocarbon pyrolysis, with implications for fuel optimization and energy applications.

Although substantial developments have been made in understanding ammonia and methane combustion, NH3/CH4 blends with the addition of alcohol additives are not well understood in terms of combustion characteristics, NOx emissions, and reaction pathways, especially at extreme conditions. It is worth noting that the presence of nitrogen in NH3 and carbon in CH4 significantly complicates the prediction and control of NOx emissions, which is highly relevant to aviation and gas turbine applications [28]. Most of the previous studies have focused on the combustion characteristics of either NH3 or CH4 individually or their interaction with conventional hydrocarbon additives; very limited attention has been devoted to the effects induced by alcohol additives-ethanol and methanol-on NOx formation and global reaction dynamics of the NH3/CH4 mixtures. Therefore, this work will fill this knowledge gap by investigating the mechanism of how the concentration of ethanol and methanol affects the NH3/CH4 combustion and pyrolysis at elevated temperatures, such as 2,000 K and 3,000 K, through systematic studies. Complex reaction networks, decomposition pathways, and the NOx formation mechanism in this work will be investigated using the ReaxFF molecular dynamics (MD) simulation method. The wide range of blends of alcohol in atomistic interactions help shed light on the synergy between NH3/CH4 and alcohol additives.

## 2. Methods and Verifications
### 2.1 Reactive force field (ReaxFF)

ReaxFF molecular simulation methodologies provide a powerful simulation framework to represent the chemical reactions at atomic-scale resolutions by dynamic accounting for bond-breaking and bond formation [29, 47]. Unlike other conventional force-field methods, where the connectivity of atoms is predefined, interactions in ReaxFF are managed through bond ordering, which enables the natural development of chemical activity in time in highly reactive material systems. This technique bridges the gap between the quantum mechanical computations, which are precise but computationally expensive, and classical molecular dynamics simulations, which are effective but non-reactive. ReaxFF allows us to model massive systems under extreme pressures and temperatures with a high degree of accuracy while undergoing complicated chemical reactions. It has been widely applied in combustion studies, yielding a wealth of information on reaction mechanisms such as the formation of intermediate species and product distributions [30]. The ReaxFF can be expressed by a function of the bond order:

$$E_{system} = E_{bond} + E_{over} + E_{under} + E_{lp} + E_{angle} + E_{tors} + E_{vdw} + E_{coul} \qquad (1)$$

where the total potential energy ($E_{system}$) in ReaxFF is given as the sum of various contributions to bond energy ($E_{bond}$), over- and under-coordination penalties ($E_{over}$ and $E_{under}$), lone-pair stabilization ($E_{lp}$), valence and torsional angle energies ($E_{angle}$ and $E_{tors}$), and non-bonded interactions including Coulombic and van der Waals ($E_{coul}$ and $E_{vdw}$) [31-32]. This form of representation of ReaxFF allows it to capture even the most complex couplings between the bonded and nonbonded interactions in a reaction system with good accuracy. Based on the parameterization of such energy terms using quantum mechanical calculations, the ReaxFF method is capable of efficiently simulating the chemical process at large systems with accurate chemical reactions.

**2.2 MD simulations**

This study involved MD simulations with the LAMMPS (Large-scale Atomic/Molecular Massively Parallel Simulator) software [38,39], an increasingly popular simulator of reactive systems. The initial configurations were generated using PACKMOL [40], ensuring a uniform distribution of molecules in the simulation box while maintaining a consistent density of 0.34 g/cm³ by changing the length of simulation box. The ReaxFF was employed to model the C/H/N/O system, allowing for the dynamic tracking of bond formation and breaking during simulations. The parameter set introduced by Zhang et al. [34] was selected, based on its superior performance in capturing the ammonia-methane reactivity and intermediate species formation. A bond order cutoff of 0.3 was used to track chemically significant interactions, and charge equilibration was handled using charge equilibration method, which was applied at every timestep to maintain proper charge distribution throughout the simulation [41]. The simulation workflow consisted of three main stages: equilibration, heating, and production. The initial configuration was first stabilized through energy minimization to eliminate the unphysical atomic overlaps. Following this, the system was equilibrated at 300 K for 100ps using the canonical NVT ensemble. A timestep of 0.1fs was applied during equilibration to resolve the system's initial dynamics while maintaining numerical stability. After equilibration, the system underwent a gradual heating phase, where the temperature was increased from 300 K to 2,000 K or 3,000 K at a rate of 20 K/ps using the NVT ensemble. This controlled ramping process mimicked experimental thermal gradients and allowed for the observation of intermediate species formation. To maintain stability during this phase, the timestep was reduced to 0.05fs accounting for the increased atomic motion and reactivity as the temperature rose [32]. The heating stage ensured a smooth transition to high-temperature conditions without introducing instabilities. The final phase of the simulation involved the production run, conducted at 2,000 K or 3,000 K for 500ps to capture the steady-state reactive dynamics. The NVT ensemble was again employed to maintain temperature control, and the timestep was further reduced to 0.01fs to accurately resolve the fast chemical reactions and molecular interactions occurring at elevated temperatures [37]. To maintain stability throughout the simulation stages, the timestep was progressively reduced: 0.1fs for equilibration, 0.05fs for heating, and 0.01fs for production. This adaptive approach allowed for the accurate resolution of fast atomic motions and chemical transformations, ensuring both numerical stability and physical reliability across all stages of the simulation. Reaction events, bond tracking, and species evolution were monitored using the REAXC package in LAMMPS, with a bond order cutoff of 0.3 to record significant chemical interactions. Post-simulation analysis was conducted using Chemical Trajectory AnalYzer (ChemTraYzer) scripts [42] and an in-house Python-based code [43-44]. ChemTraYzer facilitated the identification of reaction pathways and intermediate species, while the in-house tools were used to quantify NOx formation and analyze the evolution of reactants and products. This integrated approach ensured a comprehensive understanding of the reactive dynamics and NOx formation mechanisms in ammonia-methane systems, particularly in the presence of ethanol and methanol additives. In addition, the dynamics trajectories were visualized using OVITO [45].

The study focuses on the moderately fuel-rich conditions ($\lambda$=0.7), which are particularly relevant for real-world combustion applications where perfect stoichiometry ($\lambda$=1) is difficult to achieve. These fuel-rich conditions are expected to promote the formation of intermediate species and provide valuable insights into NOx formation dynamics, which is the primary target of this study. The detailed parameters for the combustion and pyrolysis cases are provided in Supplementary Information as Tables 1 and 2, respectively, outlining the system density, temperature range, number of molecules, and cube size in each simulation. The combustion scenarios involve $CH_4$ and $NH_3$ as the primary fuels in a ratio of 1:1, with a fixed oxygen content of 378 molecules. Alcohol additives (ethanol or methanol) are introduced by partially replacing $CH_4$ and $NH_3$ in the fuel mixture, maintaining the total number of molecules at 800 to ensure consistency in system's composition. To investigate the effects of alcohol addition, 5% and 10% levels of ethanol ($C_2H_6O$) or methanol ($CH_4O$) replacement were considered. These cases are further evaluated at two temperatures, i.e., 2,000 K and 3,000 K, to assess the temperature-dependent reactivity. The detailed configurations for each combustion case are provided in Table 3. Similar to the combustion cases, the baseline setup in the pyrolysis scenarios involves only $CH_4$ and $NH_3$, while subsequent cases include ethanol or methanol at 5% and 10% replacement levels. The total number of molecules is maintained at 800 across all cases, and the temperature varies between 2,000 K and 3,000 K. Configurations for these cases are provided in Table 4. By systematically varying the temperature and alcohol concentrations, these case setups provide a comprehensive framework for evaluating the impact of alcohol additives on ammonia-methane combustion and pyrolysis. Fig. 1(a) presents a combustion setup corresponding to cases C3-C4, which includes 5% ethanol ($C_2H_6O$) as part of the fuel mixture, while Fig. 1(b) shows a pyrolysis configuration corresponding to cases P7-P8, with 5% methanol ($CH_4O$).

**2.3 ReaxFF calibration**

In this study, the ReaxFF developed by Kulkarni et al. [33] and Zhang et al. [34] were considered. The validation process for this study was conducted in three distinct steps. First, the two force fields were compared to identify the most reactive one based on simulation results. Second, the simulation outcomes of the selected force field were compared with a published study to assess the accuracy in terms of the number of molecules over time. Finally, the values of bond dissociation energy calculated using the selected force field were compared with the experimental and theoretical results to evaluate the force field's reliability in capturing key reaction energetics [35-36]. This validation ensures that the selection of a force field that provides both chemically accurate and computationally efficient results, making it suitable for detailed investigations into the combustion and pyrolysis mechanisms of $NH_3/CH_4$ systems. The analysis consists of three stoichiometric regimes: stoichiometric ($\lambda$=1), fuel-rich ($\lambda$≈0.7), and very fuel-rich ($\lambda$≈0.3) at 2,000 K and 3,000 K. The chemical reactions that control the combustion of $CH_4$ and $NH_3$ depend on the principal oxidation pathways of these reactants. Methane ($CH_4$) reacts with oxygen ($O_2$) to produce carbon dioxide ($CO_2$) and water ($H_2O$) as per the reaction: $CH_4+2O_2\rightarrow CO_2+2H_2O$. Simultaneously, ammonia ($NH_3$) undergoes combustion in the presence of oxygen to form nitrogen gas ($N_2$) and water: $4NH_3+3O_2\rightarrow 2N_2+6H_2O$. In this study, a 1:1 ratio of $CH_4$ to $NH_3$ is considered, combining these two reactions into an overall single reaction: $4CH_4+4NH_3+11O_2\rightarrow 4CO_2+14H_2O+2N_2$. The reaction pathways for $CH_4$, $NH_3$, and $O_2$ explain the formation of the primary combustion products: $CO_2$, $H_2O$, and $N_2$. The presence of intermediates such as CO, NH, and OH also provides critical insights into the reaction dynamics and the influence of stoichiometric ratios on combustion process: $CH_4\rightarrow CH_3\rightarrow CH_2\rightarrow CH\rightarrow C\rightarrow CO\rightarrow CO_2$, $NH_3\rightarrow NH_2\rightarrow NH\rightarrow N\rightarrow N_2$, $O_2\rightarrow O\rightarrow OH\rightarrow H_2O$. Three stoichiometric ratios were examined to evaluate the combustion behaviors. Stoichiometric combustion ($\lambda$=1), where oxygen is supplied in a balanced amount, is to completely oxidize the reactants. The reaction involves 50 $CH_4$, 50 $NH_3$, and 138 $O_2$ molecules. Fuel-rich combustion ($\lambda \approx 0.7$), where oxygen is limited, results in partial oxidation. This case involves 50 CH4, 50 NH3, and 97 O2 molecules. Very fuel-rich combustion ($\lambda \approx 0.3$), where oxygen availability is significantly reduced, has limited possibility to oxidize the reactants. This reaction uses 50 CH4, 50 NH3, and 46 O2 molecules. The stoichiometric configurations for these scenarios were chosen to reflect the realistic combustion conditions and to test the performance of the two force fields under both oxygen-sufficient and oxygen-limited environments. Table 5 compares the bond dissociation energy (BDE) for CH4 and NH3 of the current study with both experimental results [35–36] and data reported by Xu et al. [4]. Figs. 2 and 3 summarize the consumption of reactants and the formation of products under various conditions. Fig. 4 compares the CH4 consumption over time between the present study and data of [4] at 2,000 K and 3,000 K.

Figs. 2 and 3 summarize the consumption of reactants and the formation of products under various conditions. At the temperature of 2,000 K in Fig. 2, the ReaxFF. 2009 [34] consistently outperformed the ReaxFF. 2012 [33] in predicting the reactivity, particularly for the consumption of NH3 and the production of H2O and N2. This difference became profound under the very fuel-rich condition ($\lambda \approx 0.3$), where the ReaxFF. 2009 predicted significant reaction progress, while the ReaxFF. 2012 showed limited NH3 consumption and minimal product formation. These results highlight the ReaxFF. 2009's ability to accurately model the influence of limited oxygen availability on combustion dynamics. At 3,000 K, the reactivity from both force fields increased with the increasing of temperature. However, as shown in Fig. 3(b), the ReaxFF. 2009 still presented a better force field, where a higher degree of reactants consumption and higher production yield of reactions like H2O and N2 could be obtained. Further, the ReaxFF. 2009 was reported as more sensitive regarding changes in stoichiometric ratios and in better agreement with theoretical trends of combustion chemistry and experimental trends. Therefore, the ReaxFF. 2009 is recommended for further MD studies. On the other hand, the ReaxFF. 2012 showed inconsistent behavior in different stoichiometric conditions, specifically under ratios of $\lambda \approx 0.7$ and $\lambda \approx 0.3$. Zhang et al. [34] further showed that the suitability of ReaxFF. 2009 in NH3/CH4 combustion studies exhibited the consistency across all conditions. Success in capturing the reactivity trend, product distributions, temperature effects and correct qualification, enables it to be used with confidence in investigations relating to combustion dynamics. Based on the obtained results, therefore, the force field developed by Zhang et al. [34] is recommended for further studies dealing with the combustion of NH3/CH4 and other similar reactive systems. Due to its higher performance, it can describe the chemical behavior with higher accuracy and gaining more insight into complicated combustion mechanisms. The selection of ReaxFF. 2009 was further validated by comparing the simulation results from the present study with the those reported by Xu et al. [4]. In particular, the number of CH4 molecules was monitored for 300 ps at high temperatures of 2,000 K and 3,000 K to test the consistency and reliability of the ReaxFF. 2009. The results are shown in Fig. 4, where a comparison of predicting the methane consumption trends has been made. The CH4 consumption trends in the present study tuned very well with those presented by Xu et al. [4] for both cases: 2,000 K and 3,000 K. At 3,000 K itself, the current study (the blue curve) showed a steep fall in CH4 molecules, and its consumption was comprehensive at around 100ps. This behavior closely resembles the one reported by Xu et al. [4] (red curve), indicating that the ReaxFF.2009 can reproduce the methane oxidation kinetics at elevated temperatures with good accuracy. The good agreement of the two datasets at 3,000 K underlines the robustness of the ReaxFF.2009 in modelling highly reactive environments. The green curve in the present study at 2,000 K reflects a slower consumption rate compared to the results at 3,000 K, reflecting the reduced reactivity at lower temperatures. Such a trend is in good agreement with the work of Xu et al. [4] (the orange curve), which also reflects a gradual decline in CH4 concentration over the simulated time. However, slight differences are observed in the later stages of the simulation beyond 200ps, where the present study predicts a marginally higher residual number of CH4 molecules compared.

### 2.4 Bond dissociation energy

The bond dissociation energy (BDE) is an important parameter for assessing the reliability of the chosen ReaxFF force field for modelling the reactivity and stability of CH4 and NH3 in combustion and pyrolysis processes. BDE not only provides a quantitative estimate of the amount of energy required to deform a given chemical bond, but also calculating the reaction energetics. The ReaxFF framework calculates the bond energy ($E$) as a function of the bond order ($BO$), which represents the degree of bonding between two atoms [32, 38]:

$$E = -D_e . BO . exp\left(p_{be2} . (1 - BO^{pbe1})\right) \tag{2}$$

$$BO = exp\left(p_{bo1}\left(1 - \left(\frac{r}{r_0}\right)^{p_{bo2}}\right)\right) \tag{3}$$

The parameters $p_{bo1}$ and $p_{bo2}$ describe how the bond order decays with bond stretching or compression relative to a reference bond length $r_0$. The parameters $p_{be1}$ and $p_{be2}$ describe how the bond energy depends on the bond order. $D_e$ represents the dissociation energy for the bond. Expression in Eq. (2) ensures that energy scales with the bond order, reflecting the bond's contribution to the system's total energy.

In the present study, values of BDE calculated for CH4 and NH3 using ReaxFF. 2009 were compared with both experimental results [36-37] and data reported in the study by Xu et al. [4]. This comparison allows for a thorough assessment of the accuracy and robustness of the ReaxFF. 2009 in representing the bond-breaking processes. In the current study, the derived BDEs for CH4 and NH3 are respectively 104.35 kcal/mol and 106.09 kcal/mol. The absolute error, in terms of experimental value, is 1.75 kcal/mol for CH4 and 2.49 kcal/mol for NH3, respectively with uncertainties of roughly 1.7% and 2.4%. The results in Table 5 demonstrate that the present study achieves better agreement with experimental values of BDE than the results reported by Xu et al. [4]. The smaller absolute errors observed here underscore the improved performance of the ReaxFF. 2009 in accurately simulating the bond dissociation energetics for CH4 and NH3, suggesting that the selected force field is well-suited for modelling combustion and pyrolysis reactions involving NH3/CH4 mixtures. The nearly equivalent values of CH4 and NH3 BDE also align with their similar molecular stabilities, as reflected in their comparable consumption rates during combustion. This supports the applicability of ReaxFF. 2009 for exploring the complex reaction mechanisms and product formations in NH3/CH4 systems under extreme conditions.

### 3. Results and Discussion
### 3.1 Combustion process
#### 3.1.1 Combustion reaction pathway

Combustion reaction pathways of NOx species under various thermal conditions were analyzed for base fuels C1 and C2 at 2,000 K and 3,000 K, respectively. Figs. S1(a) and (b) display the interconnectedness and time evolution of the NOx species in the production state. Fig. S1(a) shows the NOx species pathways for C1 at 2,000 K. Seven species were found (NO, NO2, N2, HNO3, HNO2, HNO, and N#N), and the path of the transitional species N#N is pivotal to the core in supplying several pathways both to and from various NOx species. Production of HNO3 is also notably favored at 2,000 K as many paths are going to its formation. This is indicative of a relatively oxidative environment, which would favor the stabilization of NOx species as nitric acid derivatives. The reaction graph also underlines that species such as HNO and HNO2 are transient intermediates, participating in fast reactions. These intermediates contribute to a complex interplay between the molecular nitrogen (N#N) and oxidized NOx species. This is due to the presence of HNO3 as a major product under moderately high thermal conditions (e.g., 2,000 K), where oxidation processes stabilize higher-order nitrogen oxides.

In contrast, Fig. S1(b) illustrates the NOx pathways of C2 at 3,000 K. There are only six NOx species, namely, NO, NO2, N2, HNO, HNO2, and N#N. Notably, HNO3 is absent in the pathways at this elevated temperature, signifying that the high thermal conditions disrupt the stability of nitric acid and other higher-order nitrogen oxides. Instead, as shown in Fig. S1(b), NO becomes the dominant species, both in the pathway interconnections and the temporal distribution. At 3,000 K, the reaction pathways reflect a dynamic interplay characterized by high reaction rates and rapid cycling of intermediates. N#N continues to serve as the central hub, facilitating transitions between the molecular nitrogen and oxidized nitrogen species. However, the relative prominence of NO and NO2 in the pathway highlights a shift toward simpler nitrogen oxides under these conditions. The NOx pathway diagrams for systems C3-C10 (Figs. S2-S9) give key details about how ethanol and methanol addition affect the reaction behavior, as compared with baseline systems C1 and C2 in Figs. S1(a) and (b). These lines reveal the way species connectivity, reaction states and temporal features change when alcohols are added. Looking specifically at species count and type, response-arrow complexity and time steps can identify the significant patterns among these systems. Fig. S2, displaying C3 (5% ethanol at 2,000 K), indicates a smaller overall pathway complexity as compared to Fig. S1(a) (C1 at 2,000 K). Seven species (NO, NO2, HNO, HNO2, HNO3, N2 and N#N) are identified, similar to the base fuel. But NO2 and HNO3 connectivity is much less, and changes to HNO3 are confined to the previous time periods, approximately 250ps. That means that ethanol treatment slows the stabilization of higher-order nitrogen oxides while speeding up their cycle of production and use. C4 in Fig. S3 (5% ethanol at 3,000 K) shows a simpler pathway than Fig. S1(b) (C2 at 3,000 K). HNO3 is absent from the network, while NO emerges as the dominant species, with its connectivity becoming increasingly prominent. Other species, including HNO, NO2, and HNO2, have more efficient pathways, and N#N continues to function as a reaction hub. This reduction in complexity illustrates that ethanol helps favor simpler nitrogen oxides at high temperatures. Increasing the ethanol concentration to 10% at 2,000 K, as shown in Fig. S4 for C5, can stabilize the HNO3 pathway more effectively than C3. Important peaks for transitions to HNO3 occur at several timescales, including 320ps and 360ps, indicating that more ethanol encourages HNO3 to stay longer at lower temperatures. This still leaves it more complex than the base fuels, suggesting that ethanol has continued to suppress higher-order nitrogen oxide channels. In Fig. S5, which presents C6 (10% ethanol at 3,000 K), NO again dominates the flow path, as in Fig. S3. But over time, HNO shows up more frequently, whereas HNO3 does not show up at all. The network is de-centralized, and N#N remains the transitional site for reactions to NO, NO2 and HNO. Reduction of HNO3 and stabilization of simpler intermediates at this level and temperature demonstrate that ethanol has the capacity to reduce the complexity of pathways when temperature rises. Fig. S6, which depicts C7 (5% methanol at 2,000 K), displays pathways that are generally less complex than those of the baseline system in Fig. S1(a). The seven species remain the same, but NO2 connectivity is reduced, with fewer transitions to or from HNO3. The HNO3 production pathways occur earlier between 250 and 300ps, suggesting that methanol accelerates these reactions while limiting their persistence. N#N remains the central node, and the reduced pathway complexity reflects methanol's influence in suppressing higher-order oxides. At 3,000 K, Fig. S7 (C8, 5% methanol) shows a network dominated by NO, similar to the trends observed in Fig. S3. HNO3 is entirely absent from the pathways, and HNO maintains moderate activity. N#N continues its role as a central hub connecting NO, NO2, and other intermediates. Methanol's effect at high temperatures appears to simplify the reaction network significantly, focusing on the formation of simpler intermediates such as NO.

In Fig. S8, representing C9 (10% methanol at 2,000 K), the pathways display greater stabilization of NO2 and HNO3 compared to that observed in Fig. S6. The transitions to HNO3 are more prolonged, indicating stronger stability of this species at higher methanol concentrations. N#N continues to facilitate the connectivity among NO, NO2, and other species. These changes suggest that methanol at higher concentrations enhances the stabilization of higher-order oxides at lower temperatures while maintaining a streamlined network relative to base fuels. Finally, Fig. S9, which illustrates C10 (10% methanol at 3,000 K), highlights a simplified pathway dominated by NO, similar to that shown in Fig. S7. Transitions involving HNO are dense, with pronounced connectivity to NO and N#N. HNO3 is absent, and the overall network focuses on simpler intermediates. Methanol's role in suppressing higher-order nitrogen oxides and promoting simpler nitrogen intermediates is consistent across these conditions. In all systems, ethanol and methanol help reduce pathway complexity in comparison to base fuels. Ethanol systems in Figs. S2 and S4 stabilize HNO3 pathways more effectively than methanol systems in Figs. S6 and S8) at 2,000 K. Conversely, at 3,000 K, ethanol and methanol systems in Figs. S3, S5, S7 and S9 both neutralize HNO3 entirely, leaving NO as the most prevalent species. Methanol systems will largely constrict the pathways than ethanol ones, especially when operating at elevated temperatures. These observations, based on Figs. S2-S9, underscore the subtle influence of alcohol additives on NOx reaction mechanisms. Both ethanol and methanol are demonstrated to be good candidates for minimizing NOx emissions in combustion processes by reducing pathway complexity, decreasing higher-order nitrogen oxides and increasing simpler intermediates.

Fig. 5 expresses a time-domain analysis of the number of NOx molecules produced within each system for C1-C10. Ethanol and methanol additives have different trends in different conditions for ethanol systems (C3–C4: 5%, C5–C6: 10%) and for methanol systems (C7–C8: 5%, C9–C10: 10%). Each one presents a special trend that may emphasize the possible role of alcohol additives in moderating NOx emissions. There is a marked difference in the number of molecules of different NOx species reflected in Figs. 5(a) and (b) under the combustion conditions at 2,000 K and 3,000 K, respectively. This brings out a very pronounced temperature-dependent effect on the NOx formation pathway. At 2,000 K in Fig. 5(a), NO2 is the most dominant NOx species, with its number of molecules lying within the bracket of 20 and 25 molecules, while for HNO3, this occurs between 15 and 20 molecules. The presence of other NOx species such as NO, N2O, HNO, HNO2 and N#N takes place at a relatively lower order of magnitude; it vacillates within a bracket of between 0 and 5 molecules. Such a distribution indicates that the combustion process at lower temperatures favors species formation of the type of NO2 and HNO3, which require lower activation energy to form and are thermodynamically stable. It is worth noting that a greater presence of HNO3 and NO2 could also be found at 2,000 K, where the breakdown of intermediate nitrogenous compounds is relatively slower due to reduced thermal energy availability driving the endothermic dissociation reactions.

In contrast, Fig. 5(b) shows a quite different pathway for NOx at 3,000 K. NO becomes the most dominant species, whose number rises as high as 35 to 40 molecules, while HNO and HNO2 follow, which stabilize between 10 and 15 molecules and 5 to 10 molecules, respectively. On the other hand, HNO3 is considerably less formed, whose number falls below 5 molecules, and NO2 is almost negligible. This is due to the higher thermal energy at 3,000 K that accelerates the breakdown of more stable nitrogen-containing intermediates, such as NO2 and HNO3, toward simpler species, such as NO and HNO. At such high temperature, NO production is also favored via thermal dissociation routes from molecular nitrogen, N#N, and oxidation routes under these high-energy conditions. The general trend is that higher temperatures favor the formation of simpler NOx species such as NO and HNO due to higher reaction rates and available dissociation energy. However, at lower temperatures, the more complex and stable species such as NO2 and HNO3 are built up since the decomposition reactions hardly occur. This demonstrates how significant thermal energy is in controlling the reaction pathways of combustion, and further, the formation and dissociation balance concerning the NOx species.

At 2,000 K in C3, which is shown in Fig. 5(c), 5% ethanol results in a peak NO2 count of ~24 molecules at time of about 380ps, slightly below the ~25 molecules for C1 in the base fuel. HNO3 shows stabilization with a peak of ~13 molecules at about 250ps. Other species, including HNO3, NO, and N#N, remain within the 0–5 range. The results suggest ethanol's capability to moderately suppress NO2 production while promoting the formation of HNO3, indicating a shift toward more stable nitrogen oxides. As shown in Fig. 5(e), with 10% ethanol for C5, NO2 peaks at ~27 molecules at about 350ps, slightly higher than that in C3, illustrating that increasing ethanol concentration leads to more NO2 production. However, HNO3 stabilizes with two peaks (~10 molecules each) at about 320ps and 360ps, reinforcing ethanol's role in higher-order nitrogen oxide stabilization. At 3,000 K in C4, which is shown in Fig. 5(d), NO becomes the dominant species, peaking at ~38 molecules at 450ps, significantly exceeding the ~35 molecules observed in the base fuel for C2. This trend suggests that ethanol at 5% facilitates NO2's conversion to NO at elevated temperatures. HNO shows a steady increase, reaching ~15 molecules, while N#N rises to ~10 molecules. For C6 in Fig. 5(f), with 10% ethanol, NO exhibits multiple peaks (~20–23 molecules) throughout the process, but it ends at a lower range (5–10 molecules), surpassed by HNO (peaking at ~20 molecules at about 480ps). These results suggest that while ethanol maintains high reactivity at elevated temperatures, it exhibits limited efficacy in reducing the production of total NO. At 2,000 K in C7, which is shown in Fig. 5(g), 5% methanol leads to a peak $NO_2$ count of ~17 molecules at about 280ps, a markable reduction compared to the base fuel in C1 and all ethanol systems. $HNO_3$ stabilizes with a peak of ~10 molecules. Other species, including NO and $HNO_2$, remain in the 5–10 range, indicating methanol's stronger ability to suppress $NO_2$ formation. For C9 in Fig. 5(i), with 10% methanol, $NO_2$ stabilizes between ~20–25 molecules, while $HNO_3$ peaks at ~12 molecules at about 300ps. Although $NO_2$ increases over time, methanol's ability to reduce the peak values remains obvious. At 3,000 K for C8 in Fig. 5(h), NO reaches a peak of ~35 molecules at about 400ps, equalizing ethanol systems but with a smoother trend. HNO stabilizes between 5–10 molecules, and N#N exhibits a steady increase. For C10 in Fig. 5(j), with 10% methanol, NO peaks at ~30 molecules at about 280ps, which are lower than ethanol systems at similar concentrations. HNO reaches ~15 molecules at about 200ps, while N#N gradually increases to ~10 molecules, reinforcing methanol's stronger stabilization of intermediates.

### 3.1.2 Combustion reaction frequency

Reaction frequencies, as shown in Figs. 6(a–j), further emphasize the contrasting effects of methanol and ethanol. For ethanol systems in Figs. 6(c–f), corresponding to C3–C6, forward and reverse reactions involving $NO_2 \rightleftharpoons HNO_3$ dominate at 2,000 K, while $NO \rightleftharpoons HNO_2$ gains prominence at 3,000 K. Methanol systems in Figs. 6(g–j), corresponding to C7–C10, however, exhibit higher reaction frequencies for intermediate stabilization, such as $NO \rightleftharpoons HNO$ and $NO \rightleftharpoons HNO_2$. Methanol systems demonstrate consistently higher forward and reverse reaction rates, highlighting their greater ability to stabilize the intermediates and suppress the peak $NO_2$ formation.

In combustion cases, methanol additives (C7–C10) consistently outperform ethanol (C3–C6) in lowering $NO_2$ peaks. At 2,000 K, methanol reduces $NO_2$ peak temperatures and stabilizes intermediates such as $HNO_3$. Methanol stabilizes NO production at 3,000 K and facilitates easier amplification between intermediates such as HNO and N#N. Reaction frequency analysis reveals that methanol exerts greater control over the intermediate pathways, making it a more effective NOx reduction additive. Ethanol and methanol offer potential benefits in minimizing NOx emissions and stabilizing the reactive intermediates. Methanol is more effective than ethanol in reducing $NO_2$ peaks and stabilizing intermediates under all conditions, making it an even better candidate as a NOx mitigation agent. Ethanol plays a stronger role at 2,000 K, encouraging higher-order nitrogen oxides, whereas methanol provides both steady reduction of $NO_2$ and better stability of species at both low and high temperatures.

### 3.1.3 Combustion energy dynamics

The calculations of hydrogen bond energy (HBE) and average bond energy (ABE) in Figs. 7(a) and (b) not only explain the molecular stability and structure of the combustion systems but also demonstrate the advantages of ethanol and methanol additions to base fuels. These results directly address the combustion efficiency, stability of intermediates, and the reduction of harmful emissions such as NOx. HBE trends in Fig. 7(a) show that ethanol (C3–C6) and methanol (C7–C10) systems exhibit significant decreases compared with the base fuel systems (C1 and C2). For instance, HBE in C5 (10% ethanol at 2,000 K) decreases from -6849.82 kcal/mol to -7642.2 kcal/mol while it in C6 (10% ethanol at 3000 K) decreases from -2816.03 kcal/mol to -4111.1 kcal/mol. Obviously, the further reduction of HBE reflects much stronger hydrogen-bonding interactions and more stabilizations in its molecular structure when combustion takes place. This stabilization can enhance the efficiency of energy transfer within the system, reducing incomplete combustion and excessive formation of reactive NOx species. Methanol systems also exhibit reductions in HBE, such as C9 (10% methanol at 2,000 K), which decreases from -7011.13 kcal/mol to -7607.87 kcal/mol. Although methanol has a lesser effect compared to ethanol, it also contributes to a more stable combustion environment. These results indicate that stronger hydrogen bonding would reduce the reactivity of transient NOx intermediates and thus may suppress the formation of higher-order nitrogen oxides such as $HNO_3$ and $NO_2$, which become increasingly difficult to decompose via post-combustion process. Fig. 7(b) further extends the addition of alcohol that helps stabilization by showing the variation of ABE with time. Ethanol, mainly at 10%, presents the most dramatic change in C5 starting and ending values at 2,000 K with -424.715 to -424.4675 kcal/mol and for C6 at 3,000 K from -417.1425 to -432.42 kcal/mol. These ABE reductions tend to further enforce the fact that stronger and more stable molecular bonds are realized in the combustion environment. For methanol systems, a similar trend reflects (albeit less impactful than ethanol) that ABE in C9 (10% methanol at 2,000 K) decreases from -382.7625 kcal/mol to -384.2737 kcal/mol. The reduction in ABE demonstrates that the addition of alcohols, mainly ethanol, helps stabilize the intermediates that would lead to more complete combustion and lower emissions. Higher average bond energies tend to reduce the excessive fragmentation and uncontrolled reaction pathways for a more controlled and efficient combustion process.

These results have extensive implications: it means that ethanol and methanol induce increased molecular stability and improved hydrogen bonding, which reduces the reactivity of precursors and intermediates of NOx, tackling the NOx emission problem directly. Consequently, ethanol tends to be superior, demonstrating a more obvious impact on HBE and ABE; it is a more potent additive in accomplishing cleaner combustion. The trend of methanol contribution in this direction was less pronounced, though methanol also took part in the stabilizing system and reducing emission. These results tend to point toward a path through which the improvement in combustion efficiency might be achieved with reduction in NOx emission, a sort of dual advantage with environmental and operation benefits. Their incorporation, especially ethanol, therefore, presents a viable path toward developing a more sustainable formulation of fuel. Improved hydrogen bonding interactions and increased average bond energies contribute to creating a better combustion environment, so that NOx emissions into the atmosphere may be reduced. Energy efficiency also therefore becomes enriched.

The trends in charge equilibration energy, as shown in Fig. 8(a), reveal the distinct effects of ethanol and methanol additives at varying concentrations and temperatures on combustion energetics. In the absence of alcohol additives, charge equilibration energy for C1 begins at 29,391 kcal/mol and increases modestly to 29,684.2 kcal/mol at the end of the simulation. At the higher temperature of 3,000 K, C2 starts at a slightly lower value of 26,013.7 kcal/mol but exhibits a more significant rise, reaching 30,762.7 kcal/mol. This trend suggests that temperature plays a significant role in increasing the effectiveness for the redistribution of electronic charge to stabilize the combustion environment [22]. These effects become significantly heightened upon the introduction of ethanol, more specifically at 10% concentrations. At 3,000 K, the C6 ethanol systems exhibit the greatest final charge equilibration energy, with values topping off at 35,358.4 kcal/mol. This large leap exemplifies ethanol's capability to exponentially rise and distribute electrons to effectively stabilize the charged intermediates during combustion. Methanol systems also show an increase in charge equilibration energy compared to the base fuels. In any case, their impact is less pronounced as compared to ethanol. For example, C10 (10% methanol at 3,000 K) concludes with a charge equilibration energy of 32,937.7 kcal/mol, which, while significant, remains below that of the corresponding ethanol system, C6. These results imply ethanol's better capability of modifying the electronic interactions in the combustion system, especially under higher temperatures. Also, the trend of the Coulomb energy in Fig. 8(b) further provides the stabilization influence of alcohol additives on charged species. For the base fuels, C1 exhibits a starting and ending Coulomb energy of -48,179.1 and -48,598.8 kcal/mol, respectively, and only a very limited decrease therein. While the temperature in C1 is higher, the drop of Coulomb energy is more radical in C2, decreasing from -44,584.5 kcal/mol to -53,163.5 kcal/mol. The large drop corresponds to the better stabilization of charged intermediates with increased thermal energy, thereby minimizing the runaway reactions and making the combustion processes more controllable [22].

The enhanced stabilization by the addition of alcohol can be reflected by the more significant decrease in Coulomb energies of both ethanol and methanol systems. The ethanol systems always presented a higher decrement in Coulomb energy than the methanol systems did. Precisely, C6, which corresponds to 10% ethanol at 3,000 K, reaches a minimum value in Coulomb energy of -60,675.1 kcal/mol, showing that ethanol is more favorable for increasing the intensity of the Coulombic interactions between charged species. Methanol systems are impressive but less powerful compared to those in C6. For example, C10, 10% methanol at 3,000 K, settles at -55,030.5 kcal/mol, reflecting that methanol provides much less charge stabilization as compared to ethanol under similar concentration and temperature conditions. Thus, the observation is also coherent with the observed trend in the case of HBE and ABE as obtained in Figs. 7(a) and (b). The enhanced charge equilibration and reduced Coulomb energy in ethanol systems correspond to stronger hydrogen bonding and more stable average molecular bonds. Ethanol's ability to redistribute the electronic charges aligns with its greater reductions in HBE and ABE, suggesting that it fosters a combustion environment characterized by stronger intermolecular interactions and greater stability of intermediates. In chemistry point of view, the increase in charge equilibration energy and decrease in Coulomb energy reflect the enhanced stability of charged intermediates, such as NOx species, during combustion [22-23]. Stronger Coulombic interactions reduce the fragmentation of intermediates, limiting the formation of higher-order NOx compounds such as HNO3 and NO2. Additionally, the increased charge redistribution in ethanol systems facilitates the suppression of runaway reaction pathways, contributing to more efficient and controlled combustion.

The observed trends underscore the potential benefits of incorporating ethanol as a combustion additive, particularly at higher concentrations and elevated temperatures. Ethanol's superior performance in enhancing charge equilibration and reducing Coulomb energy can be attributed to its molecular structure, which includes an additional -CH2 group compared with methanol. This molecular structural difference not only increases ethanol's potential for stronger intermolecular interactions but also enhances its capacity to redistribute the electronic charges more effectively within the combustion environment. The presence of the extra -CH2 group in ethanol contributes to greater polarization effects, which stabilize the reactive intermediates and suppress excessive fragmentation during the combustion process. Methanol, while effective in improving the charge equilibration and reducing Coulomb energy, demonstrates less impact. Its simpler molecular structure lacks the additional -CH2 group that confers ethanol with its superior properties. Consequently, methanol may be less effective in promoting the robust stabilization of charged species and facilitating the formation of stronger intermolecular bonds. This difference in molecular complexity explains ethanol's higher efficacy in mitigating the formation of NOx species and promoting complete combustion.

### 3.1.4 NOx trends and reduction mechanisms

The trends in NOx production for combustion cases C1 to C10 at temperatures of 2,000 K and 3,000 K have been shown in Fig. 9, revealing that how ethanol and methanol additives affect emissions. NOx emissions show variable behavior at 2,000 K. At 500ps the base fuel (C1) produces a NOx level of 90.79ppm; conversely, C5 including 10% ethanol obtains the lowest NOx concentration at 88.31ppm. At 100.76ppm and 110.54ppm respectively, C3 with 5% ethanol and C7 with 5% methanol produce higher NOx emissions; C9 with 10% methanol produces 94.46ppm. Using 10% methanol in fuel results in lower NOx emissions compared to the base fuel although NOx emissions rise with other alcohol additives. The variations observed appear to result from alcohol-induced changes in combustion temperature and the reaction intermediates and oxidation pathways at lower temperatures. The addition of alcohol results in a clear reduction of NOx emissions at 3,000 K. The base fuel (C2) produces 109.95ppm of NOx at 500ps while C6 with 10% ethanol results in the lowest NOx level at 66.46ppm, which is followed by C10 with 10% methanol at 76.84ppm, then C4 containing 5% ethanol at 81.63ppm, and C8 with 5% methanol at 89.99ppm. Alcohol-derived radicals such as OH become active at high temperatures to interact with nitrogen-containing intermediates, thereby stopping the creation routes of NOx [24-26]. Higher temperatures accelerate the thermal breakdown of nitrogenous species, hence lowering NOx emissions. The impact of alcohol additives indicates clear dependence on temperature levels in NOx production. While varying alcohol and concentration affects NOx emissions at 2,000 K, all additives show a continuous reduction in NOx emissions at 3,000 K.

### 3.2 Pyrolysis process
### 3.2.1 Pyrolysis reaction pathway

In pyrolysis process, which does not provide an oxidizing environment, reactions between ammonia (NH3) and methane (CH4) are mainly driven by thermal energy, and they are not oxidative reactions. This lack of oxygen changes the chemistry of the reaction completely since no oxidative intermediates such as H2O or CO2 are generated. Rather, it is dominated by the generation and cycling of nitrogen radicals (N, H2N, H5N2) and carbon intermediates (CH2O, CH3, C2H4). These decomposition processes illustrate how high temperatures can dissolve the chemical bonds, generate reactive intermediates and regulate the species stability, resulting in different reaction networks at 2,000 K and 3,000 K respectively. Oxygenation reduces the availability of simpler molecules and radicals, further highlighting the distinct nature of pyrolysis compared with combustion process. Fig. S10 presents the pyrolysis reaction pathways for the base fuels, specifically system P1 at 2,000 K in Fig. S10(a) and system P2 at 3,000 K in Fig. S10(b). These pathway visualizations offer a clear perspective on the different reaction dynamics and intermediate radical production under different temperatures during pyrolysis. In system P1 at 2,000 K, the reaction network is moderately complex, involving the key species such as CH7N, N, H5N2, H2N, CH3, H, H4N and H6N2. Nitrogen-containing molecules such as N and H6N2 are the most common of these, which serve as key nodes allowing links to other intermediates. Stability of nitrogen radicals is a result of lower thermal energy at 2,000 K, which favors the creation and stabilization of intermediate species. These pathways to CH7N and H5N2 further indicate that decomposition reactions take place predominantly in intermediates containing nitrogen. The limited production of carbon radicals, including CH3, illustrates the dim presence of carbon interactions at this lower temperature. Essentially, the network structure is an intermediate between the radical stabilization and slow intermediate decomposition. Conversely, system P2 at 3,000 K exhibits a significantly improved reaction network characterized by greater pathway complexity. Increased thermal energy enhances the decomposition of ammonia and methane, leading to the production of advanced intermediates including CH5N, CH6N, CH2N, and C2H5, along with the nitrogen-based radicals identified in P1. The significance of CHN, N#N and C2H4 as key intermediates highlights the transition to highly reactive species at 3,000 K. The lack of higher-order stabilized species such as CH7N underscores the destabilizing effects of elevated temperatures, which promotes a rapid cycling of intermediates and favors the formation of simpler radicals. The connectivity of N#N as a central hub linking nitrogen- and carbon-based intermediates is significant, highlighting its essential role in enabling transitions between these species under pyrolysis conditions. The comparison of these two systems indicates that the pyrolysis pathways exhibit distinct trends that are dependent on temperature. At 2,000 K, the reaction network exhibits enhanced stability, primarily characterized by nitrogen-based intermediates. In contrast, at 3,000 K, the elevated energy levels facilitate an increasing reactivity, resulting in a wider distribution of both nitrogen- and carbon-based radicals. This shift indicates improved decomposition dynamics and accelerated cycling of intermediates at higher temperatures, essential for comprehending the impact of thermal energy on pyrolysis reactions.

Figs. S(11–18) illustrate how alcohol additives, specifically ethanol and methanol, affect the pyrolysis pathways at temperatures of 2,000 K and 3,000 K. These findings provide valuable insights into the formation of intermediates and radicals, as well as the influence of these additives on the reaction pathways and stabilization. Fig. S11 highlights the effects of ethanol at a moderate temperature in system P3 (5% ethanol at 2,000 K). The pyrolysis pathway reveals that ethanol promotes the formation of intermediate species such as C2H6O and C2H9NO while also stabilizing nitrogenous intermediates such as H4N and H6N2. The lack of oxygen restricts the oxidation reactions, directing the pathway towards nitrogen and carbon-hydrogen intermediates. CH7N emerges as a significant nitrogen-containing intermediate that benefits from the interaction between the base fuel and ethanol. Fig. S12 depicts system P4 (5% ethanol at 3,000 K), where the pathway accelerates and is dominated by smaller radicals such as H2N and H4N. Ethanol decomposes quickly into simpler intermediates, with N#N serving as a critical node for nitrogen-based reactions. In comparison to 2,000 K, this system shows increasing radicalization and faster decomposition rates due to higher thermal energy. Fig. S13 illustrates that a higher concentration of ethanol results in greater stabilization of intermediates such as H6N2 and C2H9NO in system P5 (10% ethanol at 2000 K). The reaction pathways indicate a delayed decomposition of ethanol-derived intermediates, underscoring ethanol's role in prolonging the existence of higher-order species. This stability is particularly noteworthy given the absence of oxygen, which limits further oxidation. Fig. S14 shows system P6 (10% ethanol at 3,000 K), where the reaction pathways become significantly simpler. Ethanol quickly decomposes into small intermediates such as H2N

and H4N, while C2H5 demonstrates transient stability. In contrast to lower ethanol concentrations, the system's streamlined pathways concentrate on nitrogen radicals and smaller carbon species, which is consistent with the high-temperature pyrolysis conditions. Fig. S15 illustrates system P7 (5% methanol at 2,000 K), emphasizing the role of methanol in radical formation. Key intermediates such as CH3O, CH2O and H2 are prevalent in the pathways. Methanol facilitates the quick stabilization of carbonaceous radicals such as CH3, and its breakdown leads to the creation of smaller nitrogen species, which contrasts with the stabilization effects seen with ethanol. Fig. S16 displays system P8 (5% methanol at 3,000 K), where methanol boosts the generation of smaller radicals, including CH2O, CH3O and H2. The network becomes significantly simpler, showcasing methanol's effective decomposition and its focus on crucial intermediates such as nitrogen radicals (e.g., H2N and H4N). Fig. S17 reveals a prolonged stabilization of CH3O and CH2O alongside nitrogen intermediates such as H5N2 and H4N in system P9 (10% methanol at 2000 K). The higher concentration of methanol increases the longevity of these intermediates, allowing for a more extensive exploration of reaction pathways while keeping a moderate complexity level. Fig. S18 shows that the pyrolysis pathways of methanol are notably streamlined in system P10 (10% methanol at 3,000 K). Key intermediates such as CH2O, CH3, and H2 suggest rapid radical formation and decomposition at elevated temperatures. Methanol's influence in reducing higher-order intermediates and stabilizing smaller radicals is particularly clear in this system.

    Fig. 10(a) illustrates the time evolution of the molecular quantities for essential species in the pyrolysis system P1. After 500ps, the final molecular concentrations provide critical insights into the stability and reactivity of different species. Nitrogen (N) is the most prevalent species, with a final concentration of 13 molecules, followed by H6N2 with 9 molecules. These species imply a significant function of nitrogen and its derivatives in the breakdown processes at lower temperatures. Secondary species, including H2N (three molecules) and H5N2 (two molecules), also participate in the reaction network, underscoring the formation of intricate intermediates. Furthermore, minor species such as CH3, CH7N and H4N stabilize at one molecule each, indicating the restricted radical propagation for these species. The comparatively low concentration of hydrogen (H, 2 molecules) indicates limited availability of H atoms, which may constrain the recombination events. The dynamics of these species indicate that the pathways predominantly favor the synthesis of nitrogenous compounds at 2,000 K, which are stabilized by the reduced energy environment. This indicates that decomposition reactions proceed through a controlled mechanism with limited secondary radical interactions. Fig. 10(b) depicts the progression of species in system P2 at an elevated temperature of 3,000 K. The increased temperature significantly enhances the decomposition rates, resulting in a substantially altered distribution of species. Hydrogen (H2) predominates the system with a remarkable ultimate concentration of 293 molecules, indicating that an increased hydrogen release attributes to the elevated thermal energy. This pronounced disparity relative to P1 underscores the significance of temperature in bond dissociation and the facilitation of H2 synthesis. The high production of hydrogen obtained in the systems of 3,000 K in this work agrees with the behavior already reported in the literature concerning ammonia pyrolysis [27, 48]. Hydrogen has been described as a characteristic product of ammonia decomposition, especially under the high-temperature conditions, as reviewed by Monge-Palacios et al. [48]. Indeed, their work treats ammonia as a hydrogen carrier, underlining the pyrolysis paths that provide high yields of hydrogen. Higher temperature noncatalytic pyrolysis of ammonia, therefore, favors molecular hydrogen to be the major product because of thermal decomposition of ammonia into intermediate radicals such as H2N and H prior to their final formation of H2. This property of ammonia pyrolysis puts it in a prime position for hydrogen generation in energy-based applications, with ammonia being used as a carbon-free hydrogen carrier. Nitrogen-containing compounds demonstrate varied behaviors. N#N stabilizes at 24 molecules, signifying the pre-eminence of molecular nitrogen synthesis. CHN (26 molecules) and CH3 (12 molecules) also appear as notable species, indicating enhanced production of carbon-centered radicals at elevated temperatures. Additional significant species comprise H2N and H6N2 (10 molecules each), CH5N (9 molecules), and CH6N (5 molecules), all of which highlight the increasing reactivity and intricacy of radical pathways at 3,000 K. Minor species, including C2H4 (5 molecules), CH2N (6 molecules) and CN (3 molecules), further illustrate the complex breakdown network, which is influenced by the interaction of high energy and reactive intermediates. Elevated

temperatures promote a wider diversity of species development, highlighting the fragmentation of hydrocarbons and the production of reactive nitrogen radicals. The comparison between P1 and P2 illustrates the significant influence of temperature on pyrolysis pathways. At 2,000 K, the system demonstrates a more restricted reaction network with a reduced number of secondary reactions, promoting the stabilization of nitrogenous intermediates such as N and $H_6N_2$. Conversely, the system at 3,000 K exhibits a highly dynamic network marked by significant bond cleavage, increased molecule fragmentation and the prevalence of hydrogen and nitrogen radicals. Indeed, this contrasts with P1 at much lower $H_2$ concentrations, implying significant hydrogen evolution upon rising temperatures. To further support the fact that the increased pathway complexity results are due to the possession of greater thermal energy, P2 reveals the presence of such species containing carbon and nitrogen radicals as CHN, N#N and $CH_3$.

Figs. 6(C–J) present the time evolution of impactful radicals and intermediate species in pyrolysis systems of P3-P10 under different conditions with various alcohol additives. The investigation has emphasized the crucial role that the type of alcohol (ethanol or methanol), concentration (5% or 10%) and temperature (2,000 K or 3,000 K) play in the pathways and overall reactivity of the $CH_4$ and $NH_3$ blended systems. The presence of alcohol greatly increases the diversity of radicals in systems. Ethanol systems, for instance, have higher concentration of nitrogen-containing radicals such as CHN, N#N and $H_5N_2$ at rising temperatures. These species can take part in several critical steps of the $NH_3$ and $CH_4$ decomposition mechanisms and thus allow for a more rapid rate or longer sustainability of higher radical concentrations. Methanol systems, in contrast, show a far more limited range of radicals, with quite clear dominance from $CH_3$, $CH_4O$ and $CH_2O$, suggesting that the decomposition paths in methanol favor smaller hydrogen-rich intermediates that suppress overall reactivity compared with ethanol. The lower temperature intermediates for ethanol, including $C_2H_9NO$, $C_2H_6O$ and $CH_3O$, present transient reservoirs and contribute to the reactive species. These intermediates are more pronounced in ethanol systems and point to their higher carbon and oxygen content influencing reaction pathways. Methanol systems generate smaller intermediates only, such as $CH_2O$ and $CH_4O$, with shorter lifetimes; they may not contribute effectively to the long-term radical pool. The addition of 5% ethanol to the base fuel in P3 promotes the formation of intermediates such as $C_2H_9NO$ and $C_2H_6O$, though their concentrations peak early in the process and diminish by the end of 500ps. Nitrogen-containing radicals such as $H_6N_2$ (10 molecules) and $CH_7N$ (3 molecules) remain stable, while N stabilizes at 11 molecules. These results point to ethanol's ability to facilitate the production of intermediates that briefly enhance the reactivity before decomposing into simpler products. Increasing ethanol concentration to 10% in P5 results in higher intermediates abundance such as $CH_2O$ (4 molecules) and $C_2H_6O$ (3 molecules) with moderate stability during the process. This corresponds to the increased hydrogen-rich decomposition pathways, in which ethanol has been enhanced. Methanol addition at 5% in P7 shows a simpler profile with fewer intermediates, which are dominated by $CH_4O$ (3 molecules) and $CH_2O$ (1 molecule). Nitrogen radicals such as N (8 molecules) and $H_2N$ (1 molecule) show limited variability. This indicates methanol's tendency to prioritize the hydrogenation pathways with fewer carbon-containing intermediates. At 10% methanol concentration, P9 exhibits similar trends to P7, though intermediates such as $CH_3O$ (1 molecule) and $CH_2O$ (0 molecules by the end) peak earlier in the reaction. Nitrogen radicals such as $H_2N$ (7 molecules) and N (6 molecules) show slightly higher stability compared with P7, indicating methanol's effect in promoting smaller, more transient intermediates. At 3,000 K, ethanol's influence becomes more pronounced, with intermediates such as CHN (26 molecules), $CH_5N$ (7 molecules) and $CH_2O$ (4 molecules) maintaining stability. $H_2$ dominates with 271 molecules, while N stabilizes at 9 molecules. These results suggest that ethanol at higher temperatures promotes the formation of reactive nitrogen-containing radicals and oxygenated intermediates. Increasing ethanol concentration to 10% in P6 enhances the hydrogen production ($H_2$ reaches 323 molecules) while maintaining the intermediate stability. Key species such as CHN (25 molecules), $CH_3$ (8 molecules) and CO (8 molecules) show strong peaks, indicating ethanol's ability to promote the robust reaction pathways at elevated temperatures. Methanol at 5% concentration in P8 exhibits a distinct pattern with high $H_2$ production (290 molecules) and notable intermediate concentrations, including $CH_6N$ (21 molecules), CHN (31 molecules) and $CH_2O$ (7 molecules). Methanol's simpler carbon backbone leads to smaller, more transient intermediates, though its ability to enhance the nitrogen radical pathways at 3,000 K is evident. Methanol at 10% in P10 shows the highest H2 production (347 molecules), with consistent formation of intermediates such as CHN (36 molecules), CH3 (17 molecules) and CO (9 molecules). Methanol promotes rapid decomposition pathways, favoring smaller intermediates and robust radical production.

In the earlier combustion systems (C1–C10), the study showed a clear relationship between the alcohol additives and NOx emissions. Ethanol demonstrated the potential to modulate NOx species, with NO2 being suppressed at 2,000 K while NO dominated at 3,000 K due to the thermal dissociation of nitrogen intermediates. Methanol reduced the peak values of NO2 compared with ethanol, indicating the stronger suppression of NOx pathways. Adding alcohols such as ethanol and methanol to CH4 and NH3 blended systems during pyrolysis and combustion brings both advantages and challenges. Ethanol, through its promotion of more diverse radical pathways and enhancement in stability of intermediates, is well-suited for applications, which require high reactivity and robust radical production. Methanol, on the other hand, might serve better in scenarios that put more emphasis on either cleaner combustion or controllable hydrogen production due to the presence of simpler pathways and reduced NOx formation. However, considering that NO emissions may be higher at higher temperatures, as in combustion systems, and that intermediate species are limited in methanol systems, tailored approaches based on application goals are highlighted. That radical and intermediate pathways can be modulated through alcohol additives to underline their value in optimizing energy systems, but the balance between the environmental impact and system efficiency is a guiding consideration.

**3.2.2 Formation of key pyrolysis intermediates**

Fig. 11 shows the likely C-H-O-N molecular decomposition path from pyrolysis, illustrating the interactions between the base fuel of CH4 and NH3 blend and added alcohols such as ethanol (C2H6O) or methanol (CH4O). The visualization mechanism has been divided into three main layers: reactants, intermediate radicals, and higher order radicals and intermediates. These depositions correlate the molecular pathways involved in the pyrolysis process with overall trends of alcohol additives. The results are interpreted in terms of temperature conditions at 2,000 K and 3,000 K. The methane and ammonia comprising the base fuel system predominantly originate from primary decomposition pathways during pyrolysis, acting as foundational reactants in the process. If one considers the results concerning the same system but without adding alcohol additives to its pyrolysis, the mentioned species are dominated mostly by CH7N, H6N2, H2N, H5N2, CH3, and H. Thus, these are also in good agreement with the species obtained from the temperature-dependent studies carried out in systems of P1 and P2. The notable species for P1 at 2,000 K are 1 molecule of CH7N, 9 molecules of H6N2, and 13 molecules of N, whereas for P2 at 3,000 K, hydrogen is far ahead with 293 molecules, but there are substantial contributions from a lot of advanced radicals, such as 12 molecules of CH3, 24 molecules of N#N, and 26 molecules of CHN. The absence of oxygen-containing intermediates such as H2O and CO2 in pure pyrolysis underlines the oxygen-limited conditions inherent in base fuel systems.

With the addition of alcohol additives, however, the paths of decomposition are drastically different. The additives of ethanol and methanol provide oxygen to the reaction system and can easily produce quite different intermediates and advanced radicals. For ethanol, its major decomposition products include ethylene (C2H4), acetonitrile (C2H3N), and carbon monoxide (CO). These species are stabilized under both conditions of 2,000 K and 3,000 K, with ethanol showing a tendency to enhance the stability of intermediate radicals. For instance, ethanol favors pathways toward the formation of C2H6O (ethanol derivatives) and nitrogen-based radicals such as H2N and H4N in P3 and P4 systems. On the other hand, methanol decomposition is in favor of the production of formaldehyde CH2O, the methyl radical CH3, and carbon monoxide (CO). At higher temperatures, such as 3,000 K, methanol also promotes the formation of more complex radicals, particularly the methoxy radical (CH3O), and enhances the production of radicals that lead to species such as H2 and CH2O. One important observation is that the species formed in systems containing alcohol additives differ from those in pure pyrolysis systems. Shared species such as H2N, H4N and H2 are then present in both ethanol and methanol routes and are respectively favored under conditions of 2,000 K and 3,000 K. CHN, CH2N2 and the ethyl radical (C2H5) are the more advanced radicals and intermediates, which are mainly observed at 3,000 K, representing the role of high temperature toward the formation of higher radicals. These species, not present in the base fuel systems of P1 and P2, reflect the increased reactivity due to the addition of alcohol, especially at higher temperatures. The absence of H2O and CO2 in the decomposition pathways reflects the limited availability of oxygen in the pyrolysis process. Instead, oxygen from the alcohol additives is incorporated into intermediates such as CH2O, CO, and CH3O, which stabilize their structures before progressing to the advanced radical stage. Good agreement with the present data underlines the special role of alcohol additives in changing the decomposition pathways. Compared with pure pyrolysis, alcohols introduce pathways to enhance both the formation and stabilization of radical and intermediate species, respectively, which may have implications for optimizing the pyrolysis processes in practical applications. Fig. 11 summarizes the molecular-level differences between pure pyrolysis systems and ones with alcohol additives. While the pure pyrolysis produces mostly nitrogen-based radicals and simple intermediates, the addition of alcohols introduces more diversity in the decomposition pathways via oxygen-containing routes and promotes advanced radical formation, especially at higher temperatures. This difference in the contributions of ethanol and methanol toward intermediate and advanced species hints at the possibility of tuning the fuel composition toward specific desired outcomes in pyrolysis.

### 3.2.3 Pyrolysis energy dynamics

Figs. 12(a) and (b) illustrate how alcohol additives contribute to the modification of pyrolysis process at temperatures of both 2,000 K and 3,000 K. The trends of hydrogen bond energy and average bond energy carry more information about the decomposition dynamics of fuel systems, showing a contrasting effect with combustion system, where ethanol had a stronger influence. The starting and closing values of H-bonding energy have different conformation in various systems. H-bond starts with -1443.71 kcal/mol for P1 at 2,000K and -1180.84 kcal/mol in base fuel. The energy decreases to -1344.97 kcal/mol for P1 and to -921.88 kcal/mol for P2. These data show that the higher temperature in P2 accelerates the bond dissociation and highly reduces the hydrogen bonding. The first averaged hydrogen bond energies after the addition of ethanol are -1193.44 kcal/mol for P3 and -1174.63 kcal/mol for P5 in systems at 2,000 K, while for the systems at 3,000 K, the initial values of P4 and P6 are -1055.75 kcal/mol and -972.85 kcal/mol, respectively. By the end of the pyrolysis process, the hydrogen bond energy in ethanol-enhanced systems decreases more dramatically at 3,000 K. For instance, P6 (10% ethanol) drops to -744.25 kcal/mol, indicating a significant enhancement in bond cleavage and radical formation. Methanol, on the other hand, demonstrates a stronger effect on hydrogen bond energy, particularly at 2,000 K. Systems of P7 and P9 begin at -1331.34 kcal/mol and -1303.17 kcal/mol, respectively, which are closer to the base fuel P1. However, at the end, the values of bond energy for the methanol systems of P7 and P9 are -1102.89 kcal/mol and -1274.47 kcal/mol, respectively, reflecting less deviation compared with the ethanol systems. Methanol systems of P8 and P10 also have a steep fall at 3,000 K, where P10 decreases from -1109.22 kcal/mol to -790.11 kcal/mol, meaning that methanol is more capable of facilitating radical formation with rising temperatures. These results suggest that methanol has a greater impact on hydrogen bond disruption, particularly at 2,000 K, compared with ethanol. This aligns with methanol's simpler molecular structure and higher reactivity, which promotes the bond dissociation at lower temperatures.

For the base fuels of P1 and P2, the initial average bond energies are -518.66 kcal/mol and -508.34 kcal/mol, respectively. These values do not change significantly with time, going up to -516.26 kcal/mol for P1 and -489.75 kcal/mol for P2. This reflects the stability of the base fuel in preserving approximately the same bond energy over time, while a larger reduction is obtained at 3,000 K due to higher energy available for bond dissociation. With ethanol as an additive, the initial bond energies for systems of P3 and P5 at 2,000 K are -547.82 kcal/mol and -579.16 kcal/mol, respectively. Thus, P5 has the lowest initial bond energy among all the systems when 10% ethanol is used. At the end, the bond energy remains almost constant with P5 finishing at -578.06 kcal/mol. The ethanol systems of P4 and P6 have a similar trend, starting at -562.65 kcal/mol and ending at -547.51 kcal/mol at 3,000 K. This indicates that ethanol at 2,000 K significantly enhances the bond dissociation, while at 3,000 K, it is relatively stable. The methanol systems of P7, P8, P9 and P10 have higher initial bond energy values compared with ethanol systems, especially at 2,000 K. For example, P7 starts at -529.06 kcal/mol, while P9 starts at -537.38 kcal/mol. However, over time, the bond energy decreases more noticeably in methanol systems at 3,000 K. For instance, P10 drops from -523.84 kcal/mol to -505.83 kcal/mol. These results indicate that methanol systems experience a gradual reduction in average bond energy at higher temperatures, reflecting their enhanced ability to promote bond cleavage and facilitate intermediate formation.

Comparative results show that methanol has a more pronounced effect on pyrolysis process than ethanol, particularly at 2,000 K. Methanol systems exhibit more significant reductions in both hydrogen bond energy and average bond energy, suggesting that methanol's simpler molecular structure and high reactivity enable it to facilitate the bond-breaking and radical formation more efficiently. However, at 3000 K, hydrogen bond energy decreases deeper in ethanol systems, which stipulates that with increasing temperature ethanol becomes more and more active due to its more complex decomposition pathways via C2H2 and C2H3N. In addition, methanol is more successful as a fuel promoter concerning enhanced bond dissociation and the appearance of more radials in pyrolysis systems at 2,000 K. This is a remarkable difference compared with combustion systems, where ethanol was shown to be much more effective for NOx reduction. The trend in methanol pyrolysis follows from its function in favoring high-order intermediates such as CH2O, CO and CH3, which are relevant to the kinetic decomposition mechanism, while ethanol mainly favors C2H2 and C2H3N intermediates that become relevant at higher temperatures. These results are consistent with the potential of adding alcohol to ammonia and methane blends through bond energy dynamics and the formation of intermediates, thus affecting pyrolysis process. Such contrasting effects between methanol and ethanol make a case for choosing appropriate additives regarding the operating temperatures of interest and reaction pathways.

These trends reflect that alcohol additives have a significant impact on the energy dynamics of pyrolysis, as inferred from variations in charge equilibration energy, shown in Fig. 13(a) and Coulomb energy shown in Fig. 13(b), over the course of simulations. The trends in the charge equilibration energy show a distinction between the base fuel systems and those modified by alcoholic additives; the base fuel systems exhibit a significantly higher initial equilibration energy at the beginning of the simulation with a starting value of 66,862.6 kcal/mol for P1 and 59,452.4 kcal/mol for P2, representing relative stability due to the absence of alcohol additives within the CH4 and NH3 blends. Ethanol addition initially shows moderately reduced energy values; hence, ethanol has a clear influence on the charge redistribution process, making it more prone to chemical reactions. More significantly, it occurs for methanol addition, mostly in higher concentration such as P10: the very starting point at 3,000 K starts with a low value at 53,479.9 kcal/mol. The energy values are considerably reduced at the end of the simulation to the lowest terminal value of 38,455.9 kcal/mol for P10, in contrast to 40,633.1 kcal/mol for P2. These results evidence that methanol is more capable of assisting in charge stabilization and forming stable radicals and intermediates, especially at elevated temperatures. Trends of the Coulomb energy also echo the effect from the alcohol additives. The magnitudes of Coulomb energy for the base fuel systems are the greatest at the inception of simulations, with starting values in P1 and P2 being -83,583.6 kcal/mol and -75,302.9 kcal/mol, respectively. This value has been reduced via the addition of ethanol, notably for the higher temperature systems, where the starting value of P4 (5% ethanol at 3,000 K) was -72,900.8 kcal/mol. Methanol again shows a much stronger effect, as P10 started at -70,115.6 kcal/mol. With time, all systems tend to show a decrease in Coulomb energy due to the stabilization of charges, including the creation of stable intermediates during pyrolysis. Among the systems with methanol addition, especially at 10% concentration, the declines are larger; P10 ended at -51,808.6 kcal/mol and had the minimum value among all systems. This highlights methanol's ability to lower the Coulomb interactions and stabilize the intermediates more effectively compared to ethanol. Ethanol systems, while effective, exhibit less pronounced reductions in energy, indicating a comparatively lower impact on charge and radical dynamics.

The temperature dependence of these trends is also notable. At higher temperatures such as 3,000 K, the effects of alcohol additives are significantly enhanced. The steep declines in both charge equilibration and Coulomb energy at 3,000 K indicate that elevated temperatures accelerate the decomposition of alcohols, facilitating the formation of radicals and intermediates more efficiently. Methanol's dominance at 3,000 K is particularly evident, as its simpler molecular structure and higher oxygen content enable faster radical formation and charge stabilization compared to ethanol. Overall, the addition of alcohols, particularly methanol, explicitly alters the energy dynamics of pyrolysis, with methanol demonstrating a more pronounced effect on reducing energy levels and enhancing radical and intermediate formation. These findings underscore methanol's potential as a more effective additive for improving the pyrolysis efficiency and controlling reaction pathways. Comparison with the base fuel systems further highlights the transformative role of alcohol additives, as they significantly influence the energy distributions, facilitate the charge stabilization, and promote favorable reaction mechanisms, particularly at higher temperatures.

## 4. Conclusions

This work has extensively investigated the effects of methanol and ethanol as alcohol additives on the combustion and pyrolysis processes of ammonia-methane blends using ReaxFF molecular dynamics (MD) simulations. Systems at 2,000 K and 3,000 K were analyzed, showing the various effects of alcohol blends on reaction dynamics, NOx pathways and emission profiles. The addition of alcohol significantly altered the NOx formation mechanisms under combustion conditions. At 2,000 K, methanol suppressed NO2 formation more than ethanol, while at 3,000 K, both alcohols suppressed NO formation, again with more pronounced effects from methanol. This was accompanied by the decay of nitric acid (HNO3) production from a dominant species at 2,000 K to negligible one at 3,000 K, because reaction pathways shifted toward the simpler nitrogen oxides.

A more quantitative assessment of the NOx trends confirms that, at 3,000 K, 10% ethanol suppressed NOx formation by approximately 39.5%, followed by 10% methanol at 30.1%, compared to the base fuel. At 2,000 K, the NOx reduction was considerably lower, with 10% ethanol leading to only a 2.7% decrease, while methanol blends showed slight NOx increases compared to the base fuel. These findings emphasize the temperature-dependent effects of alcohol on NOx chemistry, suggesting that alcohol-enhanced combustion provides a more effective emission control strategy at elevated temperatures. Moreover, the increased forward and reversible reaction frequencies of the alcohol-enhanced systems provided evidence of improved intermediate stabilization and reduced peaks in the generation of NO2. Such results show potential for alcohol additives to optimize the combustion processes by limiting NOx emissions and improving fuel stability. This is further supported by the analysis for the hydrogen bond energy and average bond energy, which indicated strengthened molecular stability in alcohol-blended systems along with reduction in bond energy values of up to -424.47 kcal/mol. These changes show a dual benefit of NOx emission reduction and improvement in molecular stability during combustion.

Pyrolysis analysis revealed that the addition of alcohol favors the reaction pathways to form intermediate species and higher radicals, with effective decomposition of ammonia and methane. Methanol exhibited higher activity compared with ethanol, especially at 3,000 K, where the charge equilibration energy decreased to -38,455.9 kcal/mol with hydrogen bond energy at -744.25 kcal/mol. Intermediated radicals such as CH2O, CH3, H2N and CHN emerged obviously, reflecting that complex decomposition pathways have been facilitated by alcohol additives. Stronger radical promotion by methanol indicated higher efficacy in enhancing the process of pyrolysis. In contrast, ethanol showed more moderate effects. All these findings illustrate that alcohol additives can improve the decomposition efficiency and reduce the formation of harmful intermediates, especially contributed from methanol. This outcome underlines the potential of blended ammonia and methane-alcohol fuels to enable sustainable energy transition by pointing out clear pathways to reduction in greenhouse gas emissions and NOx pollutants abatement in high-temperature applications.

Complex reaction networks during pyrolysis pose a challenge in accurate modelling. Incorporating machine learning techniques could enable the prediction of reaction classes with a wide applicability and give a basis to construct the full skeleton reaction network for any ReaxFF MD pyrolysis simulations at a large scale [43, 46, 49-50]. These limitations will be overcome in future studies that address the scaling of these results to industrial combustion systems, the use of other additives to further optimize the strategies of emission control, and the improvement of model predictions with more sophisticated computational methods.

**Acknowledgments**

A.S. would like to acknowledge the University of Derby for the PhD studentship (contract no. S&E_Engineering_0722) and the support provided.

**Author Contributions:** A.S.: Data curation; Formal analysis; Investigation; Methodology; Software; Validation; Visualization; Writing – original draft. Z.Y.: Project administration; Resources; Investigation; Supervision; Writing – review & editing. J.X.: Data curation; Formal analysis; Funding acquisition; Project administration; Investigation; Methodology; Resources; Supervision; Writing – review & editing. N.S.: Funding acquisition; Project administration.

**Competing Interest Statement:** The authors have no competing interests.

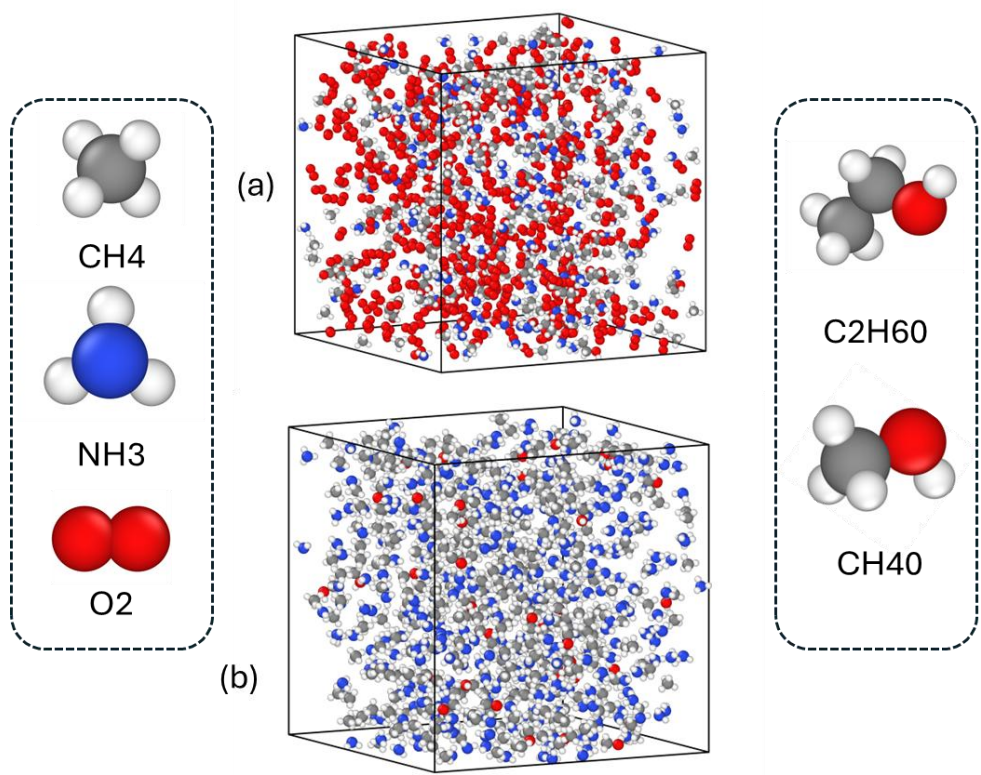

**Fig. 1.** Initial configuration of simulation systems for: (a) Combustion with 5% ethanol (C3–C4) and (b) Pyrolysis with 5% methanol (P7–P8).

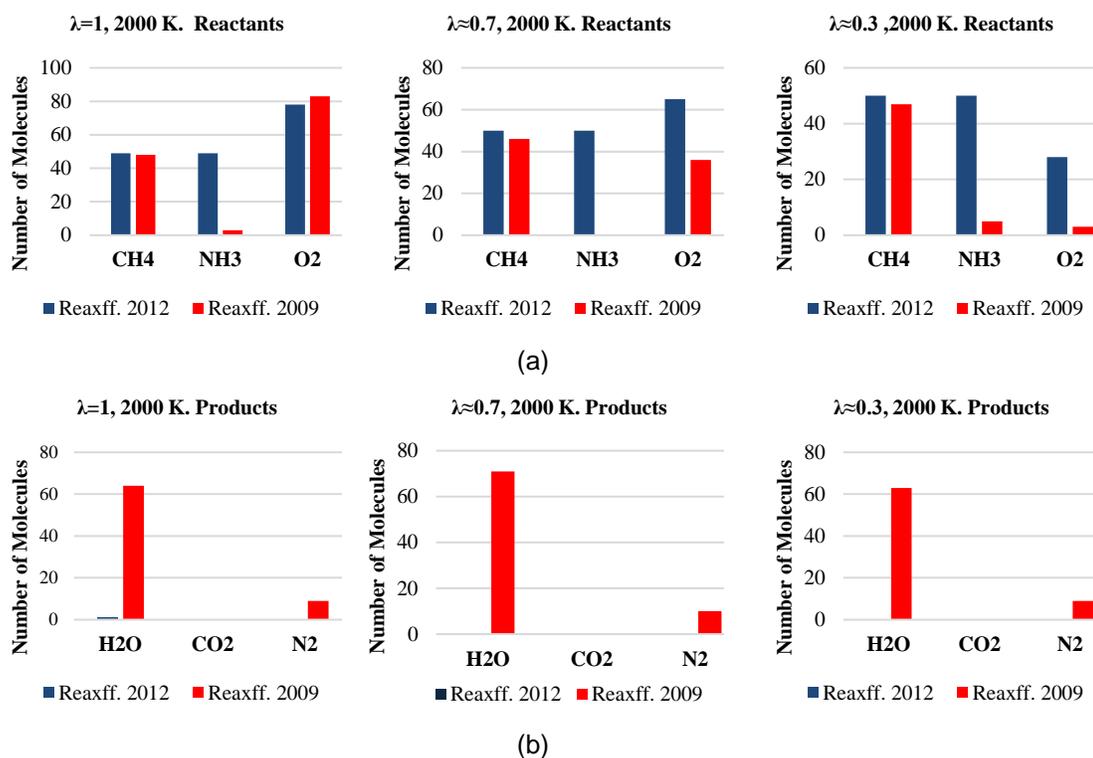

**Fig. 2.** Comparison of reactants (a) and products (b) using two ReaxFF under three stoichiometric ratios at 2,000 K.

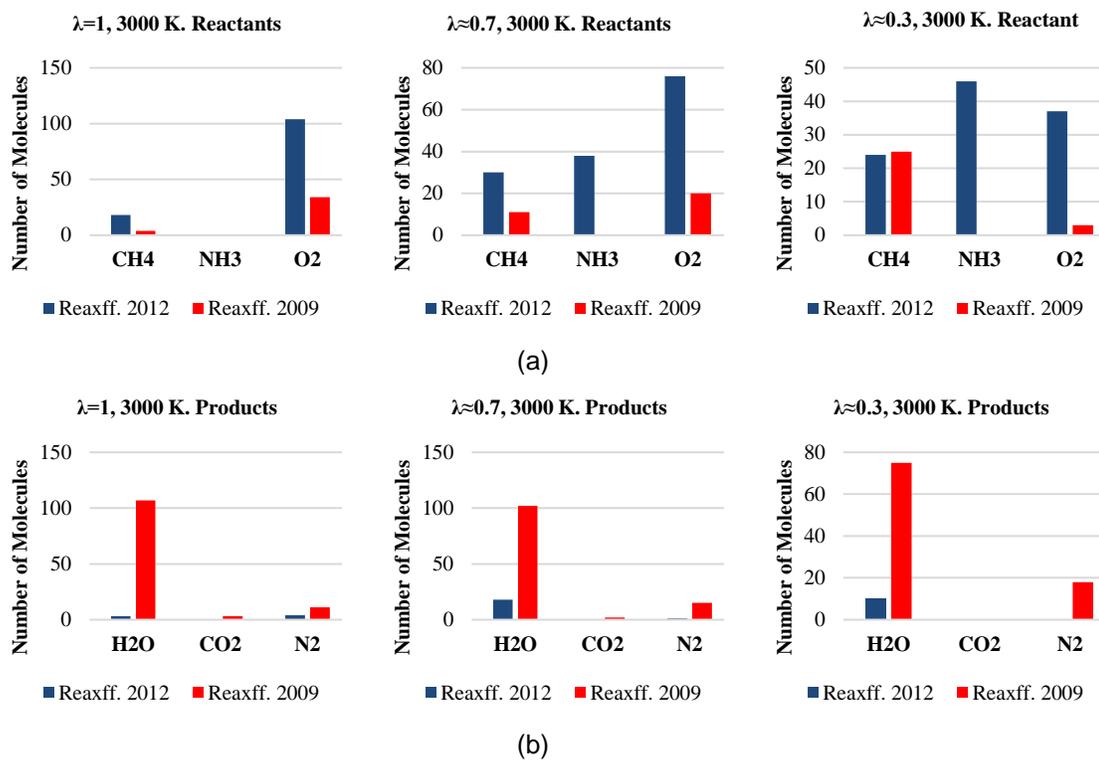

**Fig. 3.** Comparison of reactants (a) and products (b) using two ReaxFF under three stoichiometric ratios at 3,000 K.

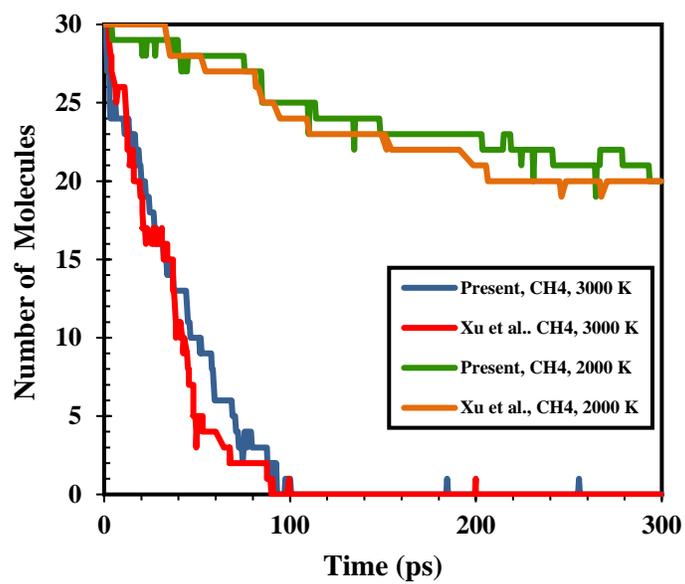

**Fig. 4.** Comparison of CH4 consumption over time between the present study and Ref. [4] at 2,000 K and 3,000 K using the ReaxFF. 2009 [34].

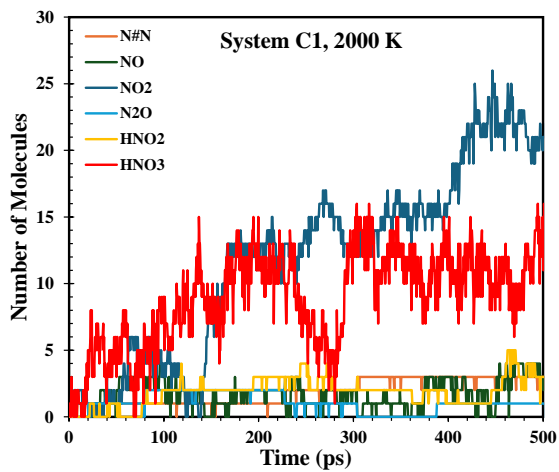
(a)

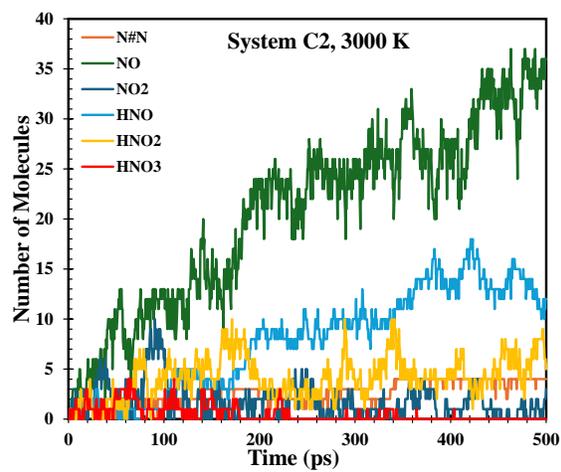
(b)

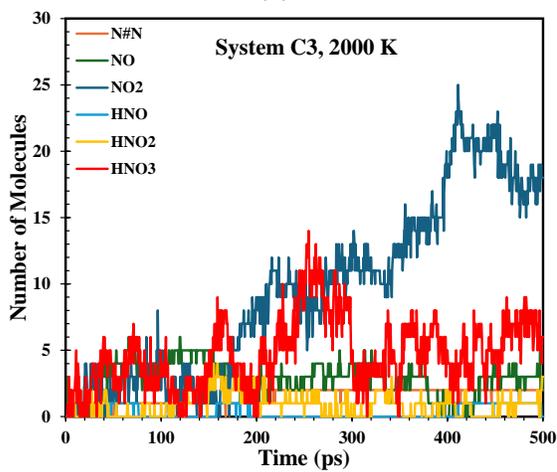
(c)

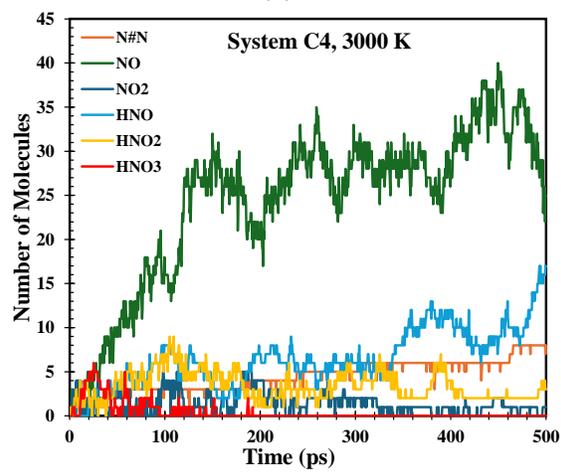
(d)

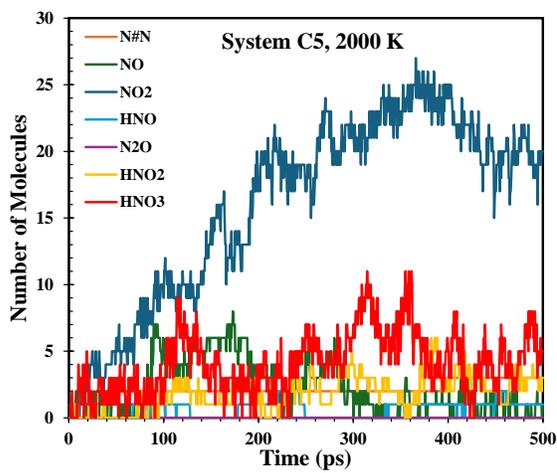
(e)

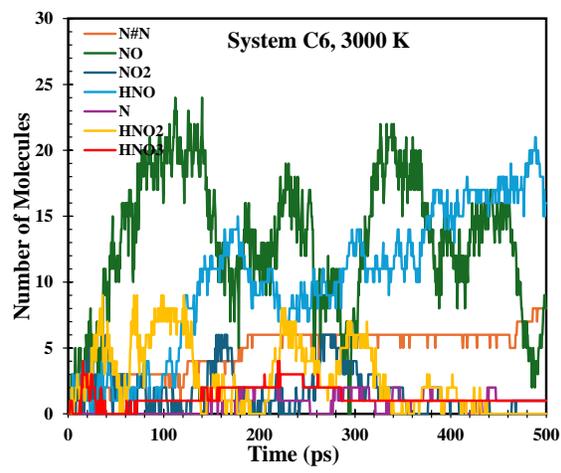
(f)

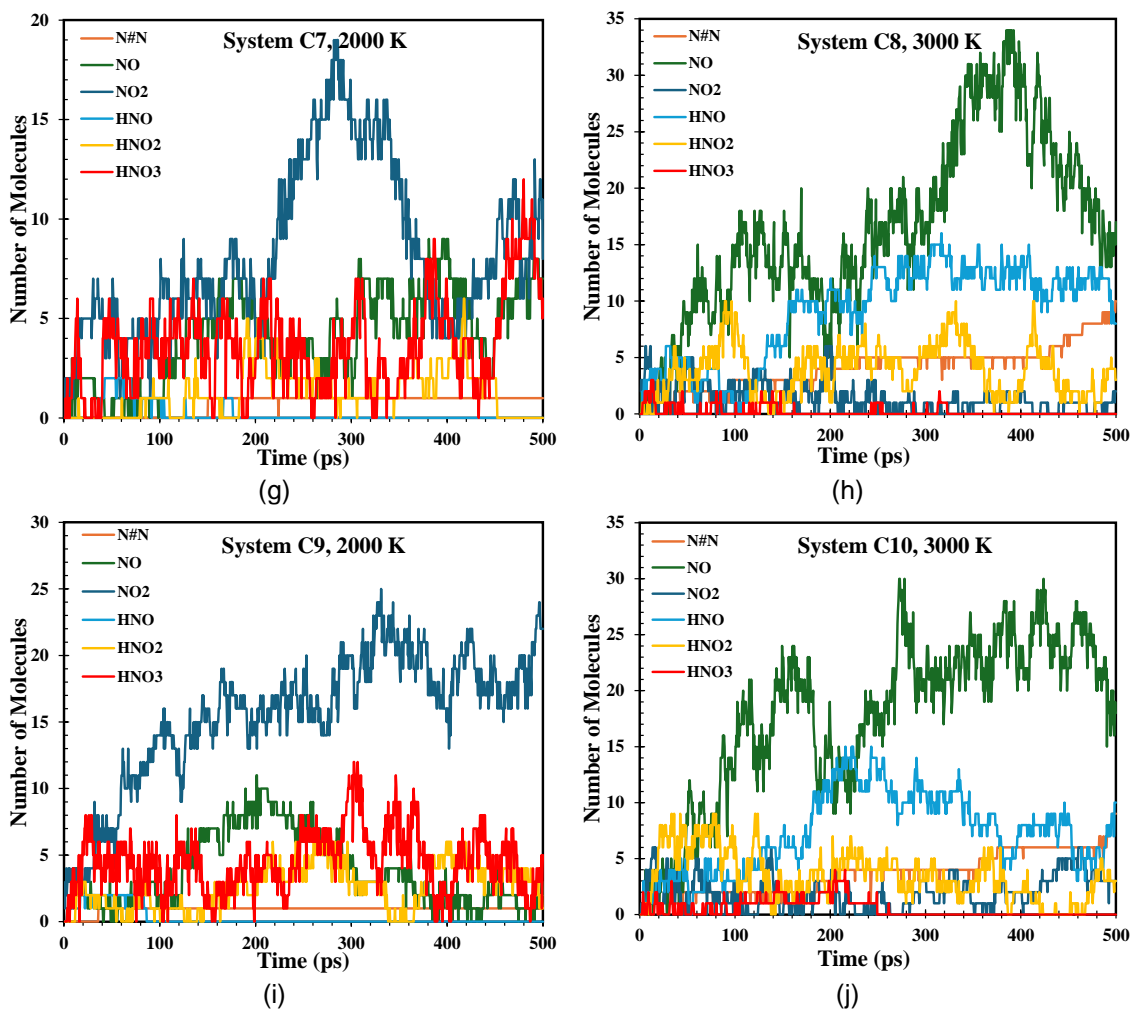

**Fig. 5.** NOx molecule counts over time during combustion process: (a) C1, (base fuel) at 2,000 K; (b) C2, (base fuel) at 3,000 K; (c) C3, (5% ethanol) at 2,000 K; (d) C4, (5% ethanol) at 3,000 K; (E) C5, (10% ethanol) at 2,000 K; (f) C6, (10% ethanol) at 3,000 K; (g) C7, (5% methanol) at 2,000 K; (h) C8, (5% methanol) at 3,000 K; (i) C9, (10% methanol) at 2,000 K; and (j) C10, (10% methanol) at 3,000 K.

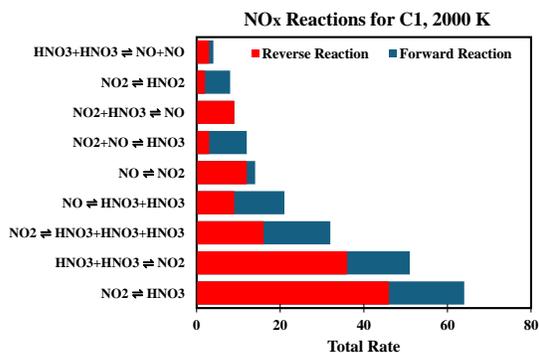
(a)

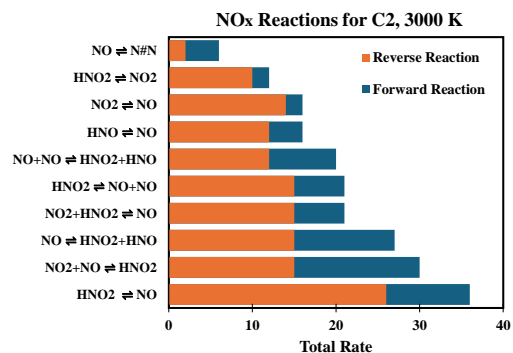
(b)

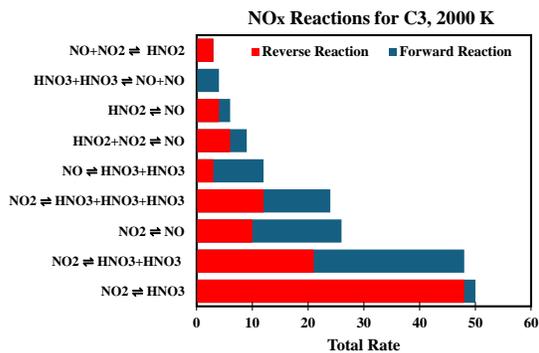
(c)

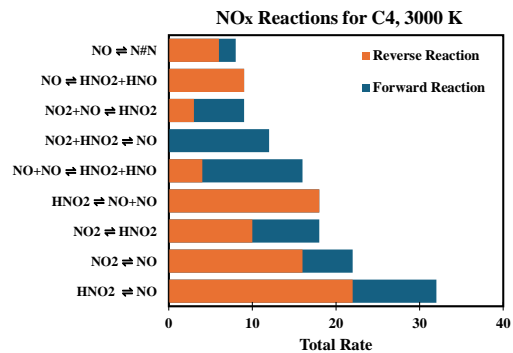
(d)

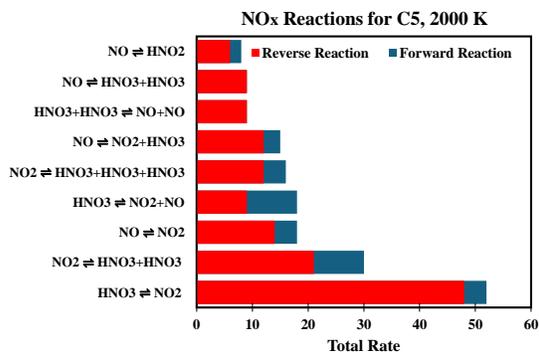
(e)

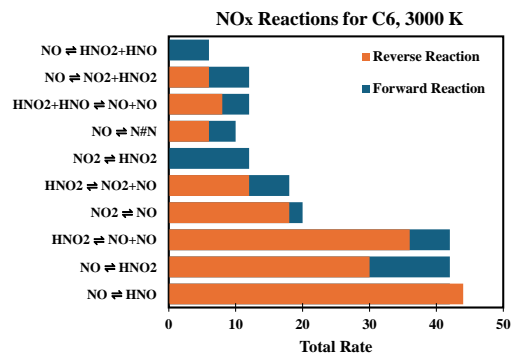
(f)

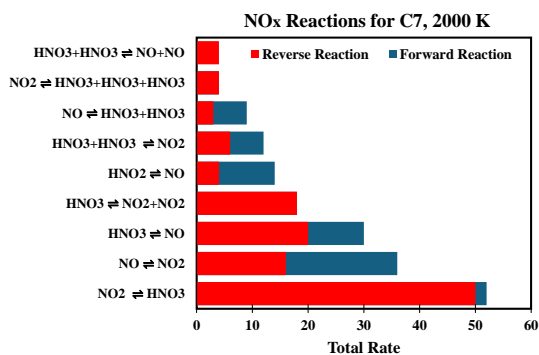
(g)

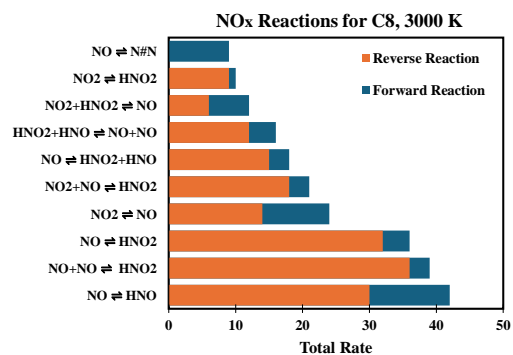
(h)

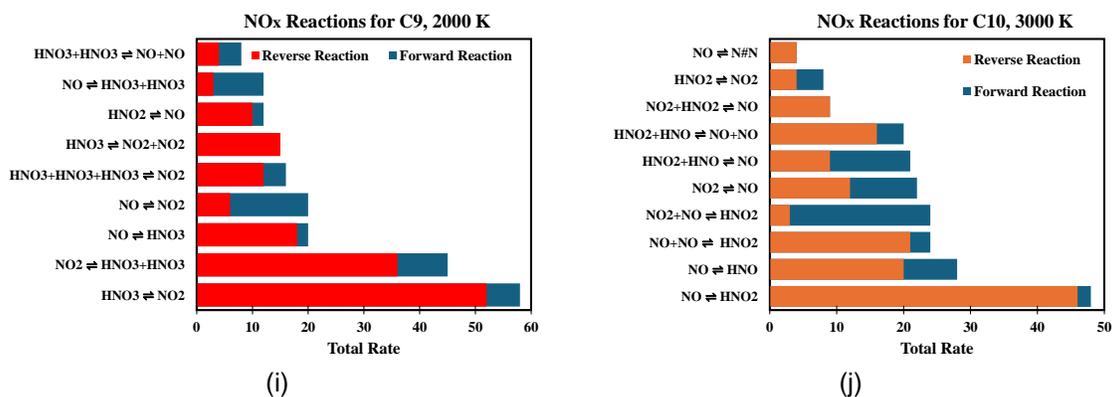

**Fig. 6.** Net reaction frequency (forward and reverse) of NOx species: (a) C1 at 2,000 K; (b) C2 at 3,000 K; (c) C3 at 2,000 K; (d) C4 at 3,000 K; (e) C5 at 2,000 K; (f) C6 at 3,000 K; (g) C7 at 2,000 K; (h) C8 at 3,000 K; (i) C9 at 2,000 K; and (j) C10 at 3,000 K.

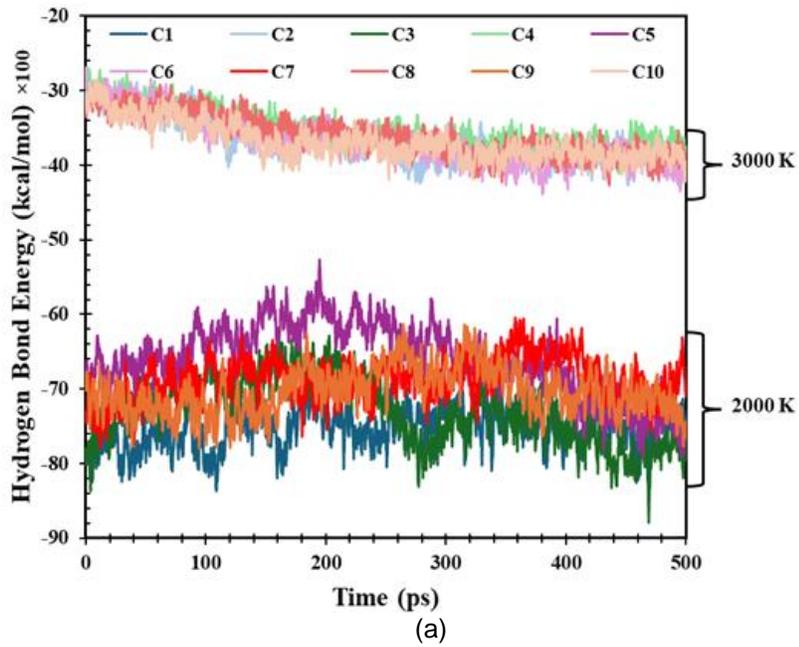

(a)

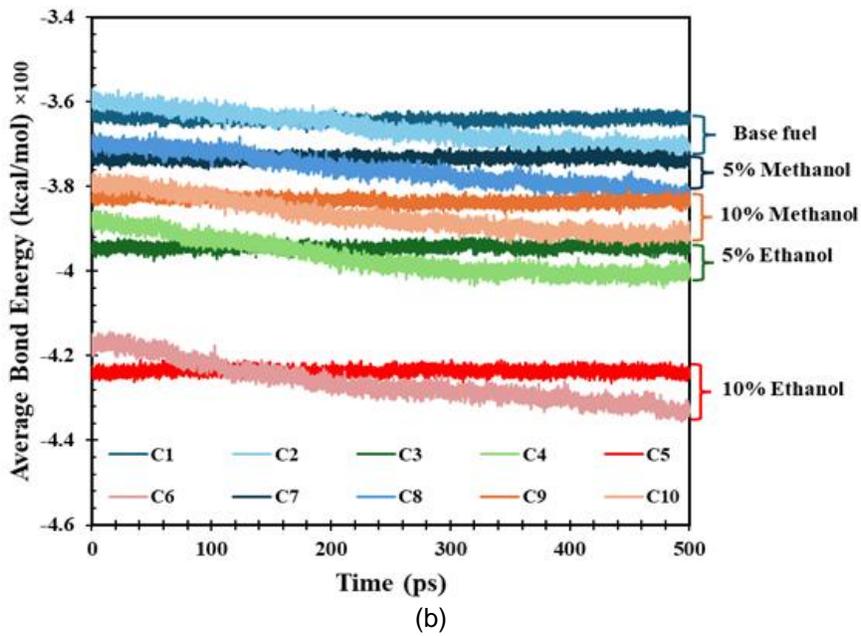

(b)

**Fig. 7.** (a) Combustion hydrogen bond energy (HBE) and (b) combustion average bond energy (ABE) trends over time for the base line and alcohol-enhanced fuels at 2,000 K and 3,000 K.

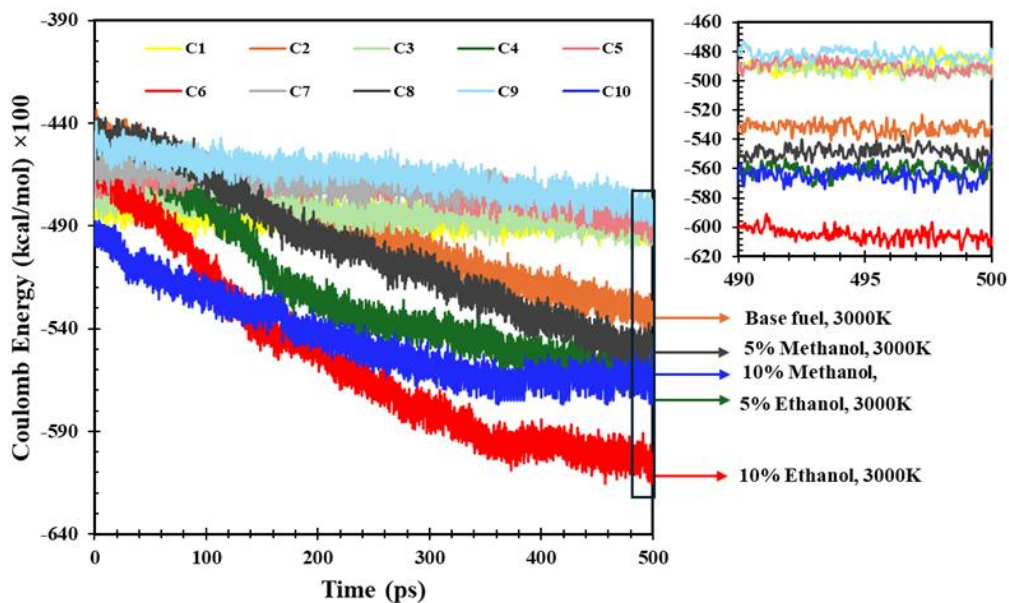
(a)

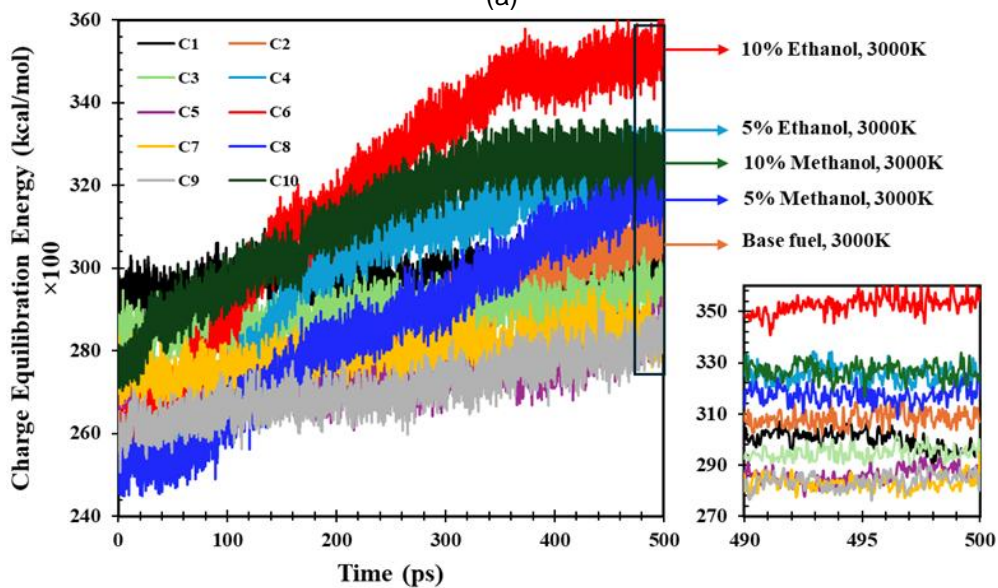
(b)

**Fig. 8.** (a) Combustion charge equilibration energy and (b) combustion Coulomb energy trends over time for the base line and alcohol-enhanced fuels at 2,000 K and 3,000 K.

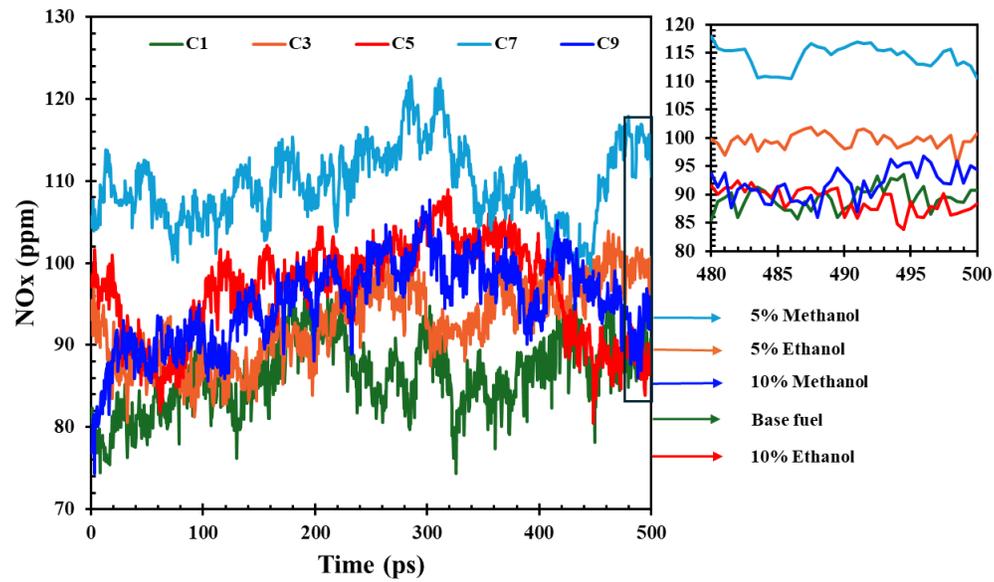

(a)

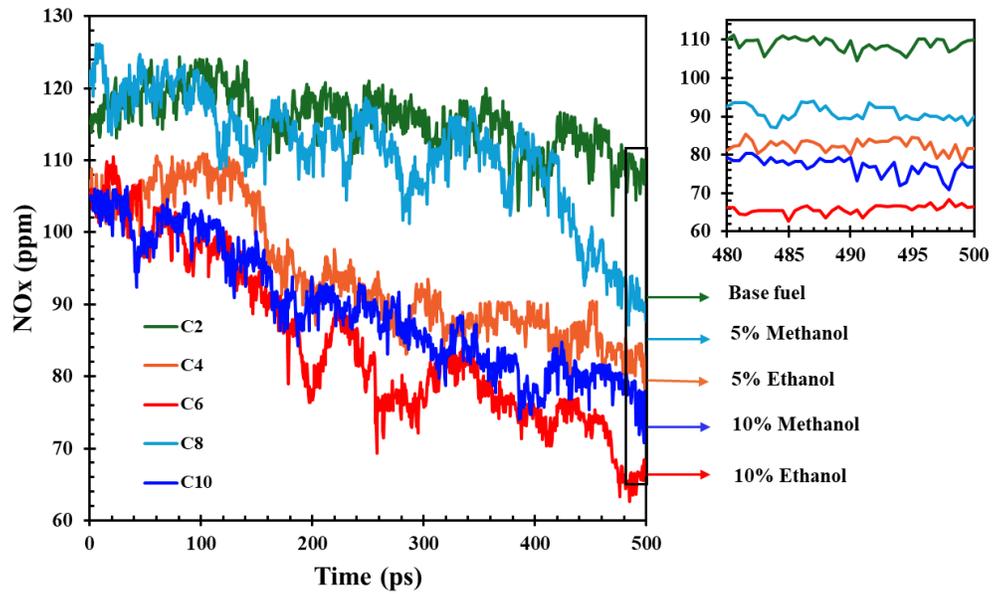

(b)

**Fig. 9.** (a) NOx formation trends over time for the base fuel and alcohol-enhanced fuels at 2,000 K and (b) NOx formation trends at 3,000 K, highlighting the reduction effect of ethanol and methanol additives.

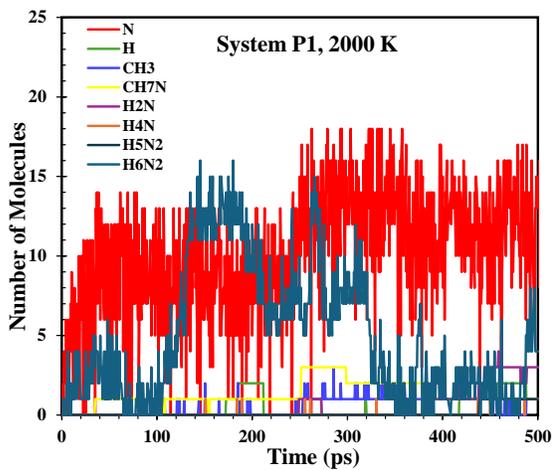
(a)

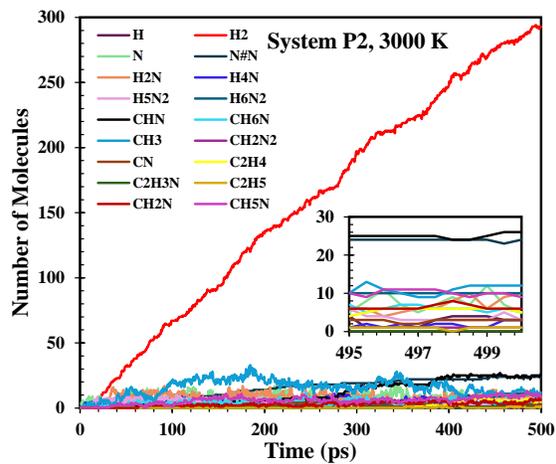
(b)

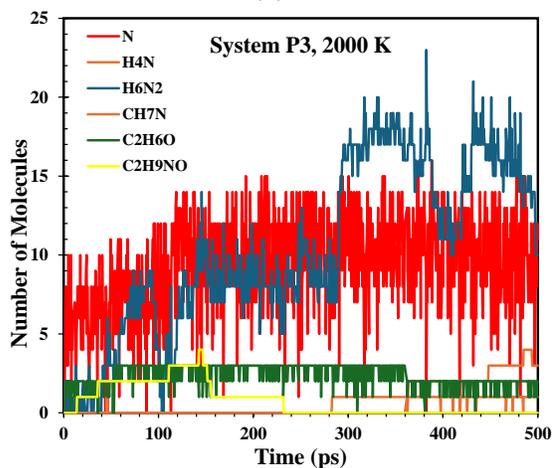
(c)

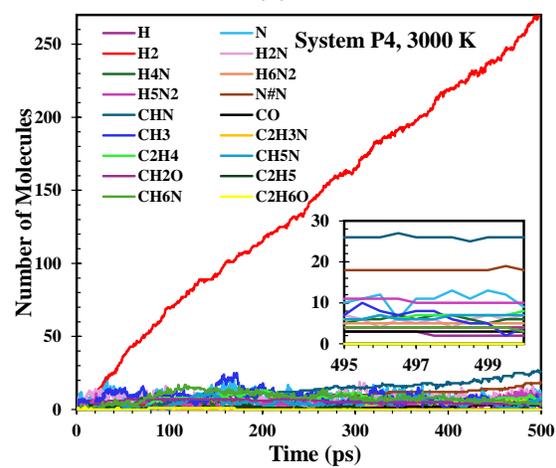
(d)

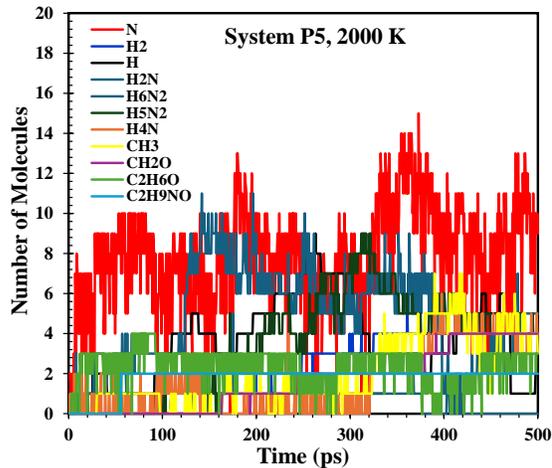
(e)

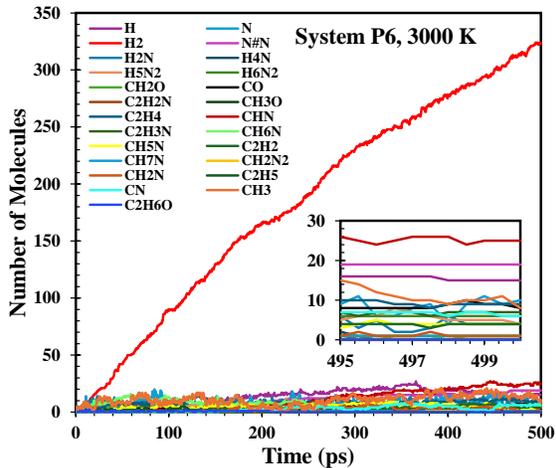
(f)

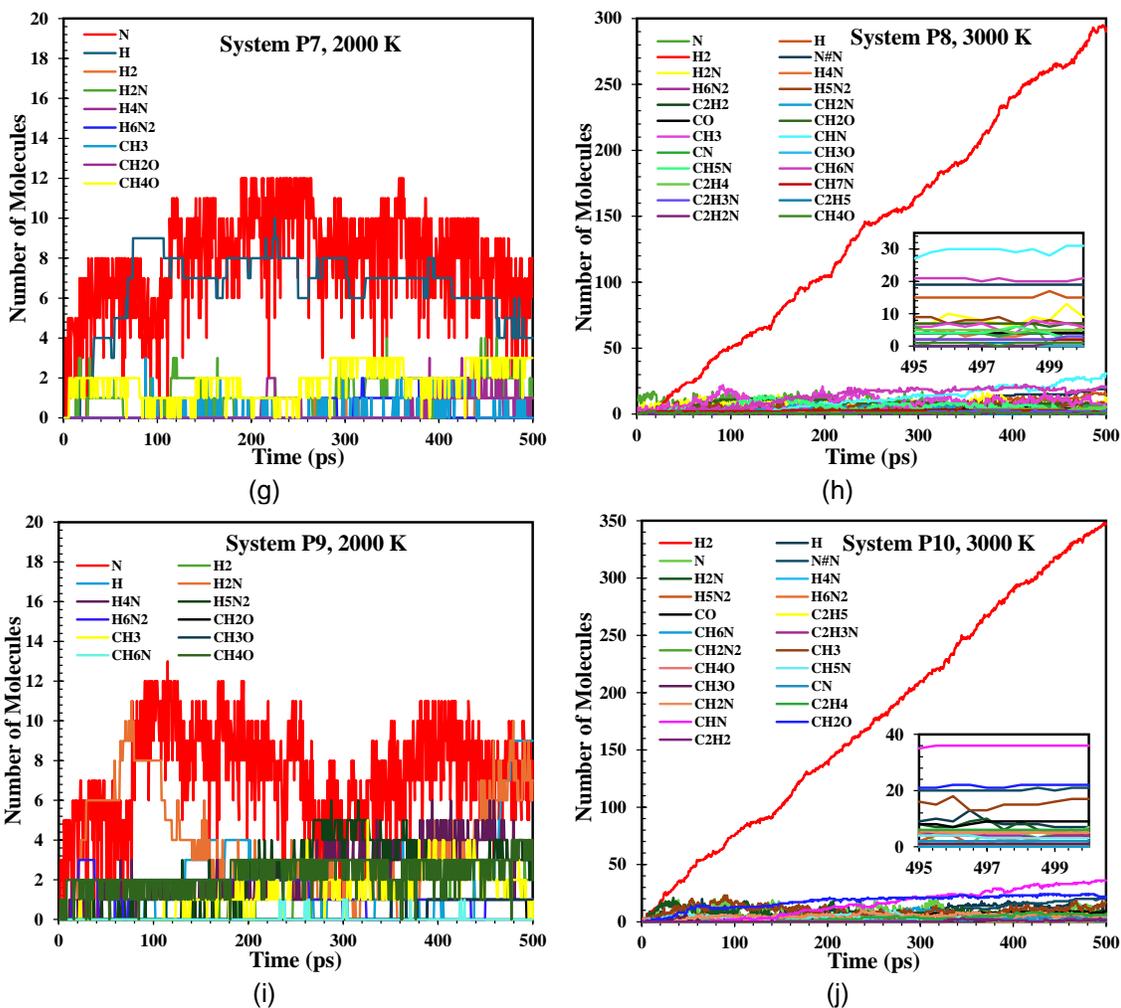

**Fig. 10.** Molecule counts over time during pyrolysis process: (a) P1 (base fuel) at 2,000 K; (b) P2 (base fuel) at 3,000 K; (c) P3 (5% ethanol) at 2,000 K; (d) P4 (5% ethanol) at 3,000 K; (e) P5 (10% ethanol) at 2,000 K; (f) P6 (10% ethanol) at 3,000 K; (g) P7 (5% methanol) at 2,000 K; (h) P8 (5% methanol) at 3,000 K; (i) P9 (10% methanol) at 2,000 K; and (j) P10 (10% methanol) at 3,000 K.

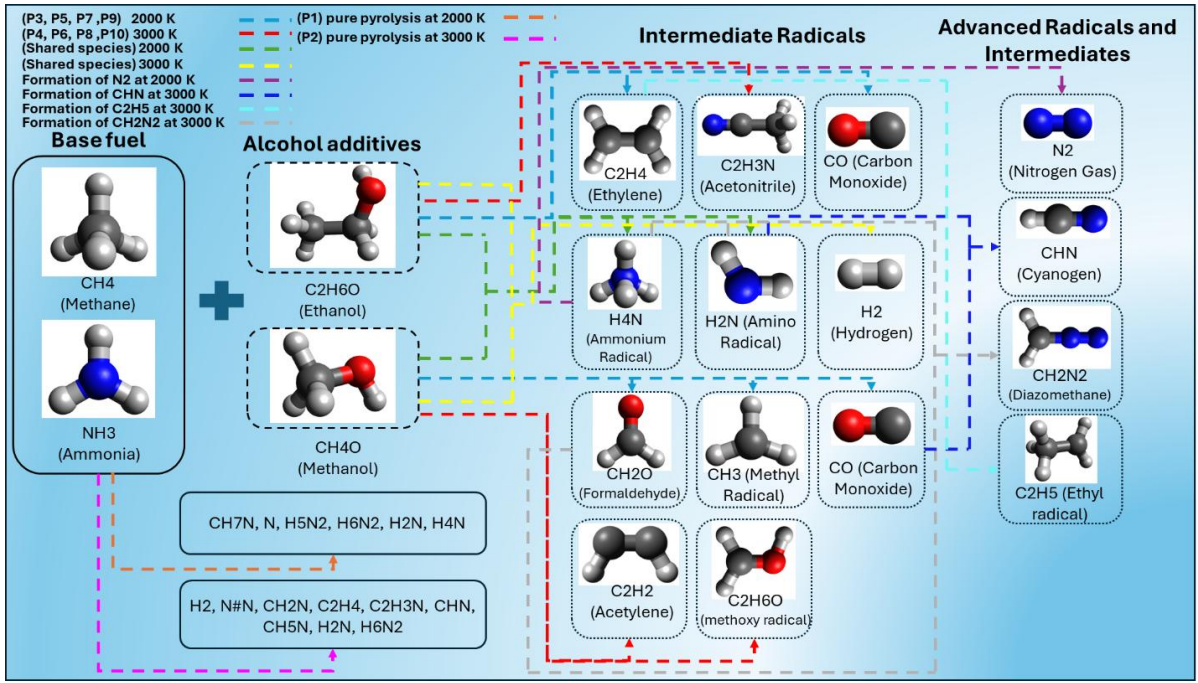

**Fig. 11.** Molecular pathways of intermediate and advanced radicals in pyrolysis with alcohol additives.

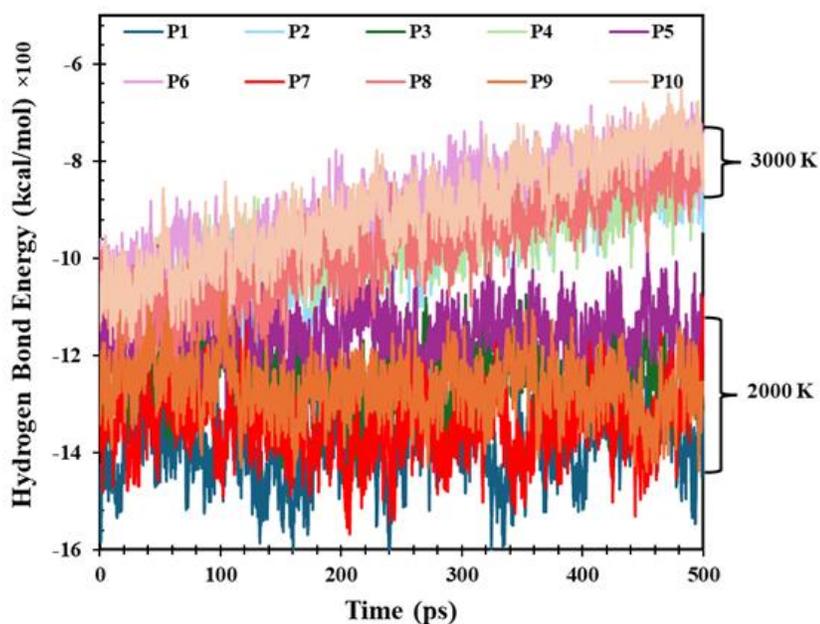

(a)

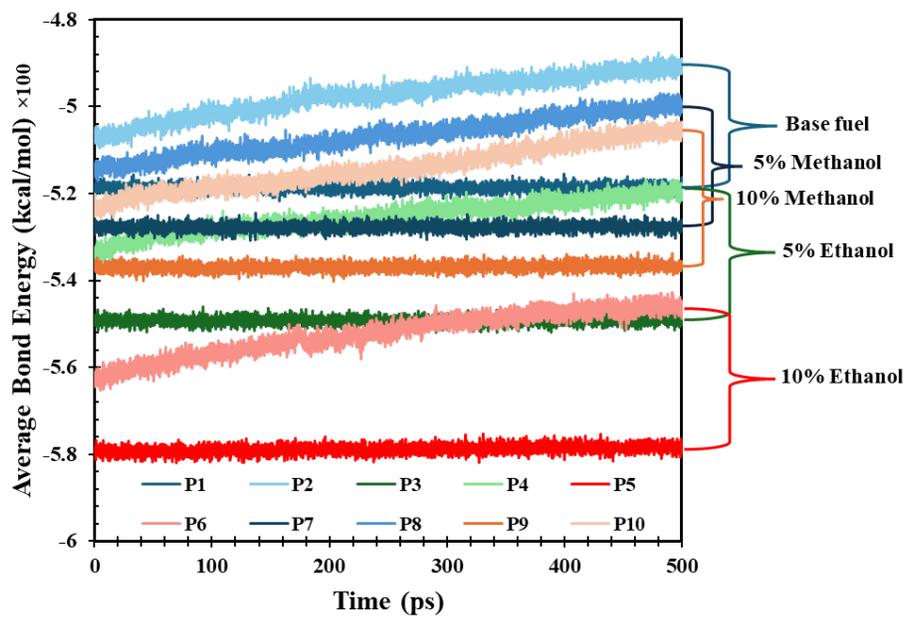

(b)

**Fig. 12.** (a) Pyrolysis hydrogen bond energy (HBE) and (b) pyrolysis average bond energy (ABE) trends over time for the base and alcohol-enhanced fuels at 2,000 K and 3,000 K.

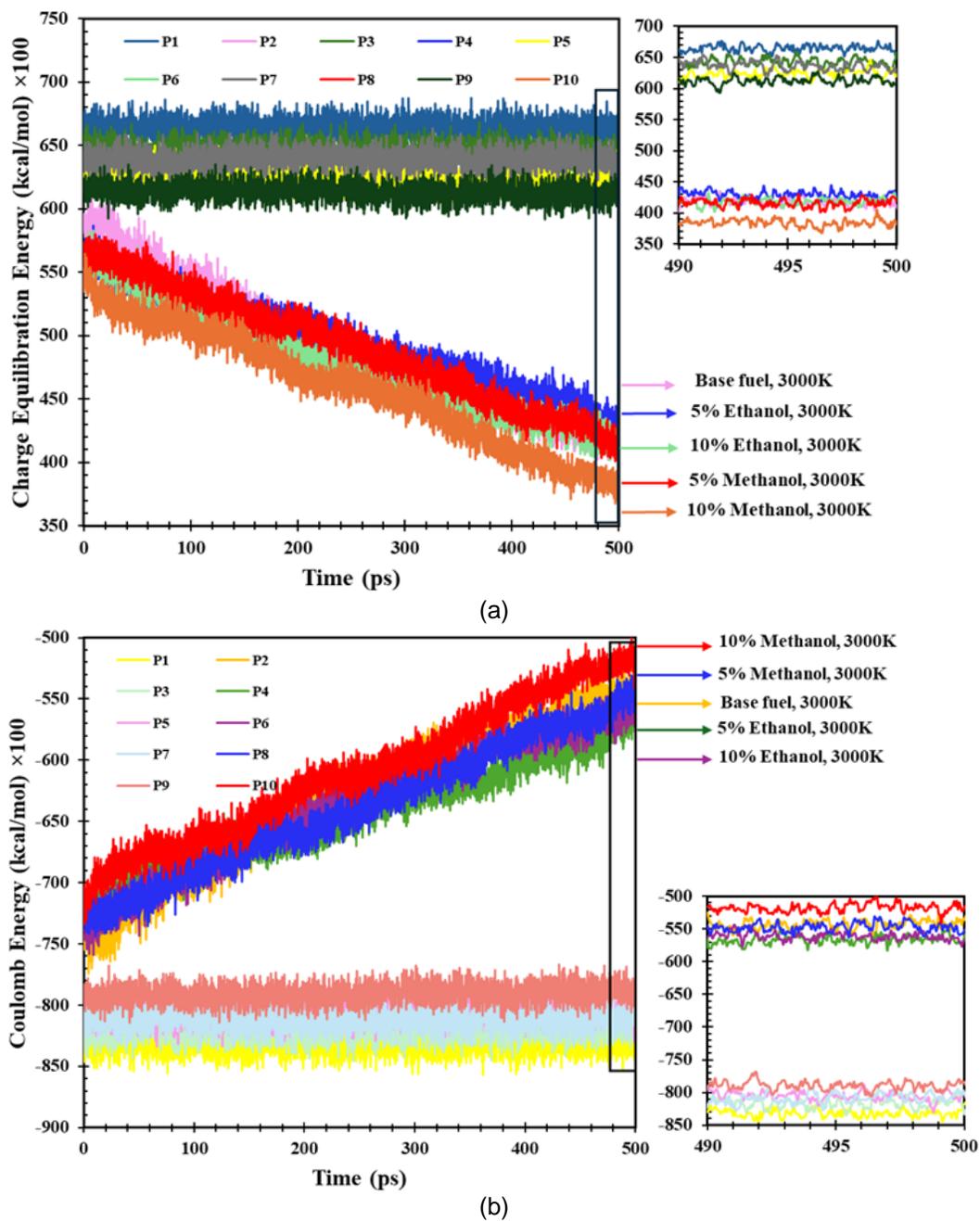

**Fig. 13.** (a) Pyrolysis charge equilibration energy and (b) pyrolysis Coulomb energy trends over time for the base and alcohol-enhanced fuels at 2,000 K and 3,000 K.

**Table 1.** Configuration and parameters in combustion cases.

| System | Density (g/cm³) | Temperature (K) | Number of molecules | Cube size (Å) |
|---|---|---|---|---|
| C1-C2 | 0.347 | 2000-3000 | 800 | 45.02 × 45.02 × 45.02 |
| C3-C4 | 0.343 | 2000-3000 | 800 | 46.11 × 46.11 × 46.11 |
| C5-C6 | 0.344 | 2000-3000 | 800 | 46.94 × 46.94 × 46.94 |
| C7-C8 | 0.344 | 2000-3000 | 800 | 45.60 × 45.60 × 45.60 |
| C9-C10 | 0.345 | 2000-3000 | 800 | 46.09 × 46.09 × 46.09 |

**Table 2.** Configuration and parameters in pyrolysis cases.

| System | Density (g/cm³) | Temperature (K) | Number of molecules | Cube size (Å) |
|---|---|---|---|---|
| P1-P2 | 0.341 | 2000-3000 | 800 | 45.12 × 45.12 × 45.12 |
| P3-P4 | 0.344 | 2000-3000 | 800 | 41.12 × 41.12 × 41.12 |
| P5-P6 | 0.344 | 2000-3000 | 800 | 42.17 × 42.17 × 42.17 |
| P7-P8 | 0.345 | 2000-3000 | 800 | 40.54 × 40.54 × 40.54 |
| P9-P10 | 0.345 | 2000-3000 | 800 | 41.12 × 41.12 × 41.12 |

**Table 3.** Combustion cases for CH4/NH3 mixtures with alcohol additives

| Case | Temperature (K) | CH4 | NH3 | O2 | Ethanol | Methanol | Total |
|---|---|---|---|---|---|---|---|
| C1 | 2000 | 211 | 211 | 378 | 0 | 0 | 800 |
| C2 | 3000 | 211 | 211 | 378 | 0 | 0 | 800 |
| C3 | 2000 | 191 | 191 | 378 | 40 | 0 | 800 |
| C4 | 3000 | 191 | 191 | 378 | 40 | 0 | 800 |
| C5 | 2000 | 171 | 171 | 378 | 80 | 0 | 800 |
| C6 | 3000 | 171 | 171 | 378 | 80 | 0 | 800 |
| C7 | 2000 | 191 | 191 | 378 | 0 | 40 | 800 |
| C8 | 3000 | 191 | 191 | 378 | 0 | 40 | 800 |
| C9 | 2000 | 171 | 171 | 378 | 0 | 80 | 800 |
| C10 | 3000 | 171 | 171 | 378 | 0 | 80 | 800 |

**Table 4.** Pyrolysis cases for CH4/NH3 mixtures with alcohol additives

| Case | Temperature (K) | CH4 | NH3 | Ethanol | Methanol | Total |
|---|---|---|---|---|---|---|
| P1 | 2000 | 400 | 400 | 0 | 0 | 800 |
| P2 | 3000 | 400 | 400 | 0 | 0 | 800 |
| P3 | 2000 | 380 | 380 | 40 | 0 | 800 |
| P4 | 3000 | 380 | 380 | 40 | 0 | 800 |
| P5 | 2000 | 360 | 360 | 80 | 0 | 800 |
| P6 | 3000 | 360 | 360 | 80 | 0 | 800 |

| P7 | 2000 | 380 | 380 | 0 | 40 | 800 |
| P8 | 3000 | 380 | 380 | 0 | 40 | 800 |
| P9 | 2000 | 360 | 360 | 0 | 80 | 800 |
| P10 | 3000 | 360 | 360 | 0 | 80 | 800 |

**Table 5.** Comparison of bond dissociation energy (BDE) for CH4 and NH3.

| Molecule | BDE process | Xu et al. (kcal/mol) | Present study (kcal/mol) | Experimental (kcal/mol) | Absolute error (present vs. exp.) (%) |
| --- | --- | --- | --- | --- | --- |
| **CH4** | 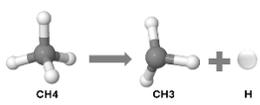 | 105.0 | 104.35 | 102.6 | 1.75 |
| **NH3** | 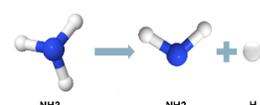 | 107.4 | 106.09 | 103.6 | 2.49 |